\documentclass[10pt,journal]{IEEEtran}
\usepackage{nccmath}
\usepackage{cases} 
\usepackage{amssymb}
\usepackage{amsmath}
\usepackage{algorithmic}
\usepackage{algorithm}
\usepackage{cite}
\usepackage{url}
\usepackage{xcolor}
\usepackage{color} 
\usepackage{cite,graphicx,amsmath,amssymb}
\usepackage{subfigure}
\usepackage{citesort}
\usepackage{fancyhdr}
\usepackage{mdwmath}
\usepackage{mdwtab}
\usepackage{caption}
\usepackage{amsthm}
\usepackage{setspace}
\usepackage{xspace}
\usepackage{graphicx}

\usepackage[acronym,nogroupskip,nonumberlist,nopostdot]{glossaries}

\DeclareMathOperator{\diag}{diag}

\newtheorem{remark}{Remark}
\newtheorem{theorem}{Theorem}

\newtheorem{lemma}{Lemma}

\newtheorem{corollary}{Corollary}

\newtheorem{proposition}{Proposition}

\allowdisplaybreaks
\makeatletter
\def\ScaleIfNeeded{%
\ifdim\Gin@nat@width>\linewidth \linewidth \else \Gin@nat@width
\fi } \makeatother

\begin{document}

\title{STAR-RIS Aided Integrated Sensing, Computing, and Communication for Internet of Robotic Things}

\author{

Haochen~Li,~\IEEEmembership{Graduate Student Member,~IEEE,}
Xidong~Mu,~\IEEEmembership{Member,~IEEE,}
Yuanwei~Liu,~\IEEEmembership{Fellow,~IEEE,}
Yue~Chen,~\IEEEmembership{Senior Member,~IEEE,}
Pan~Zhiwen,~\IEEEmembership{Member,~IEEE,}
% Xiaohu~You,~\IEEEmembership{Fellow,~IEEE}
\thanks{Copyright (c) 20xx IEEE. Personal use of this material is permitted. However, permission to use this material for any other purposes must be obtained from the IEEE by sending a request to pubs-permissions@ieee.org. (Corresponding author: Yuanwei~Liu and Pan~Zhiwen.)}
\thanks{Haochen~Li and Pan~Zhiwen are with National Mobile Communications Research Laboratory, Southeast University, Nanjing 210096, China, and also with Purple Mountain Laboratories, Nanjing 211100, China (email: lihaochen@seu.edu.cn, pzw@seu.edu.cn).}
\thanks{Xidong~Mu is with the Centre for Wireless Innovation (CWI), Queen's University Belfast, Belfast, BT3 9DT, U.K. (e-mail: x.mu@qub.ac.uk).}

\thanks{Yuanwei Liu is with the Department of Electrical and Electronic Engineering, The University of Hong Kong, Hong Kong (e-mail: yuanwei@hku.hk).}

\thanks{Yue~Chen is with the School of Electronic Engineering and Computer Science, Queen Mary University of London, London E1 4NS, U.K. (e-mail: yue.chen@qmul.ac.uk).}

}

\maketitle
\begin{abstract}
    A simultaneously transmitting and reflecting reconfigurable intelligent surface (STAR-RIS) aided integrated sensing, computing, and communication (ISCC) Internet of Robotic Things (IoRT) framework is proposed. Specifically, the full-duplex (FD) base station (BS) simultaneously receives the offloading signals from decision robots (DRs) and carries out target robot (TR) sensing. A computation rate maximization problem is formulated to optimize the sensing and receive beamformers at the BS and the STAR-RIS coefficients under the BS power constraint, the sensing signal-to-noise ratio constraint, and STAR-RIS coefficients constraints. The alternating optimization (AO) method is adopted to solve the proposed optimization problem. With fixed STAR-RIS coefficients, the sub-problem with respect to sensing and receiving beamformer at the BS is tackled with the weighted minimum mean-square error method. Given beamformers at the BS, the sub-problem with respect to STAR-RIS coefficients is tacked with the penalty method and successive convex approximation method. The overall algorithm is guaranteed to converge to at least a stationary point of the computation rate maximization problem. Our simulation results validate that the proposed STAR-RIS aided ISCC IoRT system can enhance the sum computation rate compared with the benchmark schemes.
\end{abstract}


\section{Introduction}

The Internet of Things (IoT) has catalyzed transformative advances in connectivity and automation by enabling seamless interaction between physical devices and the digital world~\cite{palattella2016internet,malik2021energy}. Within this landscape, the Internet of Robotic Things (IoRT) emerges as a specialized paradigm that leverages robotic systems to collect, transmit, and analyze data in diverse environments, ranging from terrestrial landscapes to air and water~\cite{simoens2018internet,ray2016internet}. IoRT offers unprecedented opportunities for remote sensing, exploration, and autonomous operation in environments where direct human presence is impractical or hazardous~\cite{motlagh2017uav,cui2021integrating}.

Despite its promising potential, IoRT faces multifaceted challenges that impede its widespread adoption and operational efficacy. These challenges encompass technical limitations associated with environment perception~\cite{chen2022isacot}, onboard computation~\cite{tayeb2017survey}, and communication reliability~\cite{xing2020reliability}. Addressing these challenges is imperative to unlock the full capabilities of IoRT and realize its transformative impact across various domains, including disaster response, environmental monitoring, industrial automation, and infrastructure management.

In terms of environment perception, the localization of robots plays a critical role in enabling efficient and safe navigation within IoRT systems. By continuously updating and tracking the positions of robots, the BS can dynamically plan optimal paths to avoid collisions~\cite{9369969} and optimize robot movements based on real-time environmental conditions and task requirements~\cite{9750934}. This capability is particularly crucial in scenarios where multiple robots operate concurrently in shared environments~\cite{everett1989survey}. Through active sensing, the position of the target robots can be obtained by the BS~\cite{9724202}.

% ~\cite{qiao2021survey}

Onboard computation is of vital importance for automatic robots.  With the ability to actively acquire information, the automatic robots can spontaneously make decisions and decide on the following actions. However, the decision-making process for automatic robots often entails computationally intensive tasks, posing challenges due to the limited power and central processing unit (CPU) resources of mobile robots, potentially becoming bottlenecks for instant on-board computing~\cite{9257498}. To address this limitation, the concept of mobile edge computing (MEC) has been introduced, allowing computation tasks to be offloaded to powerful MEC servers located at BSs~\cite{mao2017survey}. By leveraging MEC, automatic robots can transfer computation-intensive tasks to the MEC servers, where they are processed efficiently. {For mobile robots supported by onboard batteries, energy consumption limitations are critical. With mobile edge computing, robots can offload part of their computation to edge servers, thereby reducing their power consumption. 
}

The successful implementation of MEC relies heavily on reliable communication links between robots and BSs~\cite{sabella2016mobile}. However, challenges arise when wireless channels between the BS and robots are obstructed by static or moving obstacles, which can significantly impede the efficiency of task offloading for the robots~\cite{10032506}. Reconfigurable intelligent surfaces (RIS) offer a potential solution to this issue. Equipped with numerous reflecting elements, a RIS can dynamically adjust the propagation of wireless signals incident upon it by tuning the phase shift of each element~\cite{9140329,9847080,10550177}. Based on the aforementioned analysis, the RIS aided integrated sensing, computing, and communication (ISCC) framework is a promising architecture for IoRT. %\cite{tse2005fundamentals}

\subsection{Prior Works}
The paradigm for ISCC systems can be categorized into two main approaches: user-sensing based ISCC and BS-sensing based ISCC. In the user-sensing based ISCC approach, users are required to simultaneously carry out sensing and communication tasks (dual function)~\cite{huang2023energy,huang2023unmanned,xu2023uav,ding2022joint,9763441,huang2023mobile}. The authors in~\cite{huang2023energy} focus on jointly designing sensing and offloading beamformers for IoT users with the goal of maximizing energy efficiency within the ISCC system, which is crucial for the sustainable operation of battery-powered IoT devices. Meanwhile,~\cite{huang2023unmanned} and~\cite{xu2023uav} explore the utilization of unmanned aerial vehicles (UAVs) as dual-function users in ISCC systems. In~\cite{huang2023unmanned}, the UAV trajectory is optimized to maximize the energy and time consumption. In~\cite{xu2023uav}, the optimization of UAV trajectories and offloading beamformers is carried out with the objective of achieving a trade-off between task offloading rate and sensing beampattern gain. The above works are limited to single device cases and thus are not applicable to IoRT systems where multiple robots need service from the same BS. The authors of~\cite{ding2022joint} and~\cite{9763441} investigated the ISCC system with multiple users. In~\cite{ding2022joint}, the authors propose a novel approach using multi-objective optimization to optimize the transmit waveforms of users in an ISCC system with dual function users aiming to emulate an ideal radar beam pattern while controlling offloading energy consumption. In~\cite{9763441}, the authors focus on addressing the challenges of multi-cell interference environments in ISCC systems by solving a joint device association and subchannel assignment problem. To provide reliable and low-latency edge computing service for fault-sensitive applications, a ISCC system with short-packet transmissions is investageted in~\cite{huang2023mobile}.

For the user-sensing based ISCC, the performance of the sensing and communication tasks are constrained from three aspects. Firstly, dual-function users incur higher power consumption due to the simultaneous execution of sensing and communication tasks. This increased power demand is particularly challenging in IoRT environments where robots are often powered by onboard batteries. Secondly, the deployment of antenna arrays on mobile users is constrained by size limitations, which can restrict the sensing performance of these devices. Lastly, designing and implementing advanced hardware that supports both sensing and communication functionalities can increase complexity and cost, potentially limiting the scalability and deployment flexibility of IoRT networks. Thus, it is necessary to explore the BS-sensing based ICSC systems for IoRT.

Works~\cite{cheng2022optimized,he2023integrated,huang2022integrated,xu2023intelligent} investigated the BS-sensing based ISCC. In \cite{cheng2022optimized}, the authors employ time division multiple access (TDMA) to partition the ISCC system into distinct communication, sensing, and offloading phases within different time slots. The proposed sum communication and computation rate maximization problem is tackled with an optimized time partitioning policy. In a similar time division manner, authors in~\cite{he2023integrated} proposed a sensing-centric ISCC system where the object of sensing service is to achieve action recognition. A sensing accuracy maximization problem is solved under the communication and computation quality of service requirements. However, the limited time available for each function in traditional TDMA schemes may lead to inefficiencies in system design, motivating the need for more integrated approaches. To achieve simultaneous sensing, communication, and computation, authors in~\cite{huang2022integrated} proposed a RIS-aided ISCC system. The sum terminal energy consumption is minimized under the sensing and computation constraints. Furthermore, authors in~\cite{xu2023intelligent} proposed an ISCC system with radio-frequency-chain-free users and adopted the RIS as a passive transmitter instead of a passive relay. This scheme is promising in IoT systems where the terminal devices are cost and energy limited.
% \begin{table*}[t]
% \caption{{Notations Used in This Work}}\label{Symbols}
% \centering
% \begin{tabular}{|c||c|}
% \hline
% Symbol & Definition\\
% \hline
% $L$ & The number of DRs\\
% \hline
% $N$ & The number of STAR-RIS elements\\
% \hline
% $N_t, N_r$ & The number of BS transmit and receive antennas\\
% \hline
% $\rho_n^t, \rho_n^r, n\in\mathcal{N}$ & The reflection and transmission amplitude coefficients of the $n$-th STAR-RIS element\\
% \hline
% $\theta_n^t, \theta_n^r, n\in\mathcal{N}$ & The reflection and transmission phase coefficients of the $n$-th STAR-RIS element\\
% \hline
% $\mathbf{\Phi}_t, \mathbf{\Phi}_r$ & The reflection and transmission matrices of the STAR-RIS\\
% \hline
% $\mathbf{w}$ & The sensing beamformer\\
% \hline
% ${\mathbf{n}}$ & Noise at the BS\\
% \hline
% $\mathbf{H}$ & The channel between the BS and the STAR-RIS\\
% \hline
% $\mathbf{H}_{\mathrm{SI}}$ & The self-interference channel at the FD BS\\
% \hline
% $\mathbf{h}_{u,l}, l\in\mathcal{L}$ & The channel between the STAR-RIS and DR $l$\\
% \hline
% $P_u$ & The transmit power of DRs\\
% \hline
% $P_b$ & The power budget of the BS\\
% \hline
% $\mathbf{A}$ & The sensing channel\\
% \hline
% $\mathbf{u}_{l}, l\in\mathcal{L}$ & The receive beamformer for extracting information for DR $l$\\
% \hline
% $\mathbf{u}$ & The receive beamformer for sensing\\
% \hline
% \end{tabular}
% \end{table*}
\subsection{Motivations and Contributions}
 
{When adopting the conventional RIS to facilitate the offloading links between the robots and the BS, the robots must be situated on the same side of the RIS. This geographical restriction may not always be satisfied in real communication systems, especially for mobile robots in IoRT systems. As a remedy, the concept of simultaneously transmitting and reflecting RISs (STAR-RISs) is used to address this limitation~\cite{9690478}. Although conventional reflecting-only RISs can cover all robots when they are deployed at the end of the zone, the STAR-RISs can provide larger coverage and better channels than conventional RISs~\cite{lei2023noma,wu2021coverage}. STAR-RIS technology partitions the incoming signal into transmission and reflection regions simultaneously. This innovative approach enables full-space coverage and provides additional degrees of freedom (DoFs) for beamforming design, paving the way for more efficient and reliable communication in dynamic robotic environments. 

Given these advantages, the STAR-RIS aided ISCC framework holds promise as an architecture for IoRT.  However, the full space coverage and increased DoFs introduced by STAR-RIS also pose new challenges in system design. Firstly, there are additional optimization factors to consider, such as phase shifts and amplitudes for both transmission and reflection. Secondly, the STARS beamforming arrangement imposes additional coupling constraints, significantly increasing the design's complexity. These obstacles lead to new optimization problems compared with conventional RIS beamforming and encourage us to study STAR-RIS aided ISCC in IoRT systems. To our knowledge, there is no prior research on STAR-RIS for ISCC.}

Contributions of this paper can be summarized as follows:
\begin{itemize}
{\item A novel STAR-RIS aided ISCC IoRT framework is first proposed. This framework aims to establish reliable communication links for robots that require MEC services and enable sensing of specific robot of interest by the base station BS. The problem of maximizing the sum computation rate is formulated to optimize the beamformers at the BS and the coefficients of the STAR-RIS under constraints including the BS power constraint, the sensing signal-to-interference-plus-noise ratio (SINR) constraint, and constraints on the STAR-RIS coefficients.}

%  to establish reliable communication links for robots requiring MEC service and sense the robot that BS is interested at.

% A novel framework is introduced, termed STAR-RIS aided ISCC (Integrated Sensing-Computing-Communication) for Internet of Robotic Things (IoRT). This framework aims to establish reliable communication links for robots that require Mobile Edge Computing (MEC) services and enable sensing of specific robots of interest by the base station (BS).
% where a BS receives the offloading signals from DRs and carries out TR sensing at the same time. 

% The computation rate maximization problem is formulated to optimize the sensing and receive beamformers at the BS and the STAR-RIS coefficients under the BS power constraint, the sensing signal-to-noise ratio (SINR) constraint, and STAR-RIS coefficients constraints.

\item We first reformulate the sensing constraint in the proposed optimization problem using an equivalent communication model. Subsequently, the reformulated problem is tackled using the alternating optimization (AO) method. The sensing/receiving beamformers at the BS and the coefficients of the STAR-RIS, are optimized iteratively with each other fixed during alternating steps. With fixed STAR-RIS coefficients, the sub-problem with respect to sensing and receiving beamformer at the BS is tackled with the weighted minimum mean-square error (WMMSE) method. { Given beamformers at the BS, the rank constraint in the sub-problem with respect to STAR-RIS coefficients is tacked with the penalty based method. The resulting problem from the penalty-based approach is then solved using the successive convex approximation method.} The proposed algorithm is guaranteed to converge to at least a stationary point of the computation rate maximization problem.

\item The simulation results demonstrate that the proposed AO algorithm for solving the proposed sum computation rate maximization problem can converge. Besides, the sum computation rate depends on multiple parameters in the ISCC IoRT system, including the sensing threshold, the power available at the BS, and the number of RIS elements. Additionally, the advantage of adopting STAR-RIS in ISCC IoRT system is verified.
\end{itemize}

\subsection{Organization and Notations}
The remainder of this paper is organized as follows: Section II provides the system model for the proposed ISCC IoRT system, where the STAR-RIS model, signal model, communication model, sensing model, and computation model are introduced. Section III presents the problem formulation for the sum computation rate maximization problem and employs an AO-based approach to address the optimization problem. Section IV discusses simulation results, while Section V concludes this paper.
 
Lowercase, uppercase, and bold lowercase letters denote scalars, vectors, and matrices, respectively. The space of $M \times K$ dimensional complex matrices is denoted as $\mathbb{C}^{M \times K}$. The superscripts $(\cdot)^\mathrm{T}$, $(\cdot)^\mathrm{*}$, and $(\cdot)^\mathrm{H}$ signify transpose, conjugate, and conjugate transpose operations, respectively. The vector containing the main diagonal elements of matrix $\mathbf{A}$ is expressed as $\text{diag}(\mathbf{A})$. $\text{Bdiag}\left( \cdot \right)$ denotes the block diagonal operations. The functions $\text{tr}(\mathbf{A})$ and $\text{rank}(\mathbf{A})$ refer to the trace and rank of matrix $\mathbf{A}$. The real component of a complex number is extracted using $\Re(\cdot)$. The absolute value of a complex number is represented by $|\cdot|$. Sets are denoted using calligraphic letters, such as $\mathcal{A}$.
% The notations used in this work is summarized in Table~\ref{Symbols}.

\section{System Model} \label{sec:system_model}
\begin{figure}[!tbp]
\centering
\includegraphics[width=0.5\textwidth]{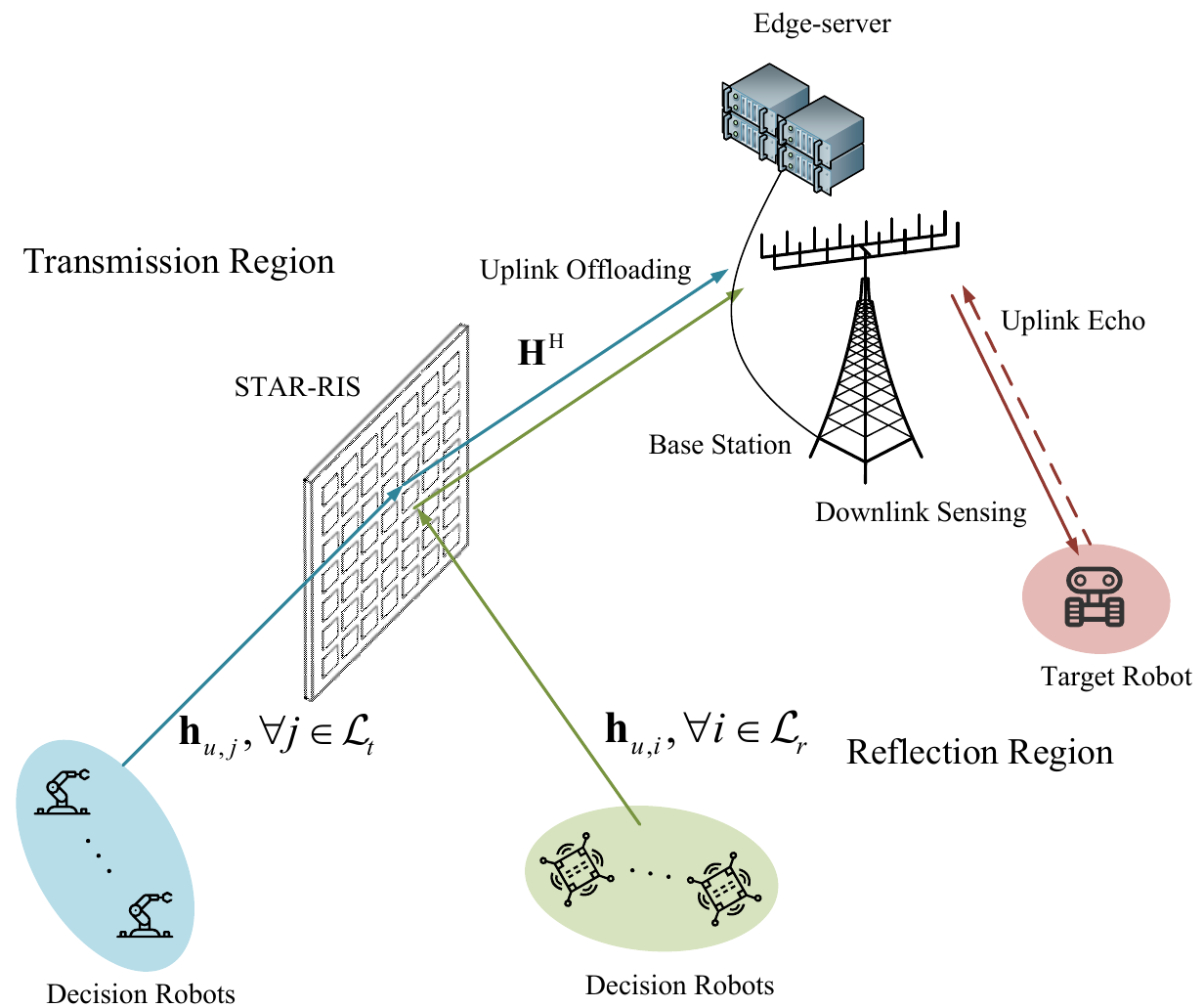}
\caption{The system model for the proposed STAR-RIS aided integrated sensing, computing, and communication system.}
\label{system_model}
\end{figure}
\begin{figure}[!tbp]
\centering
\includegraphics[width=0.5\textwidth]{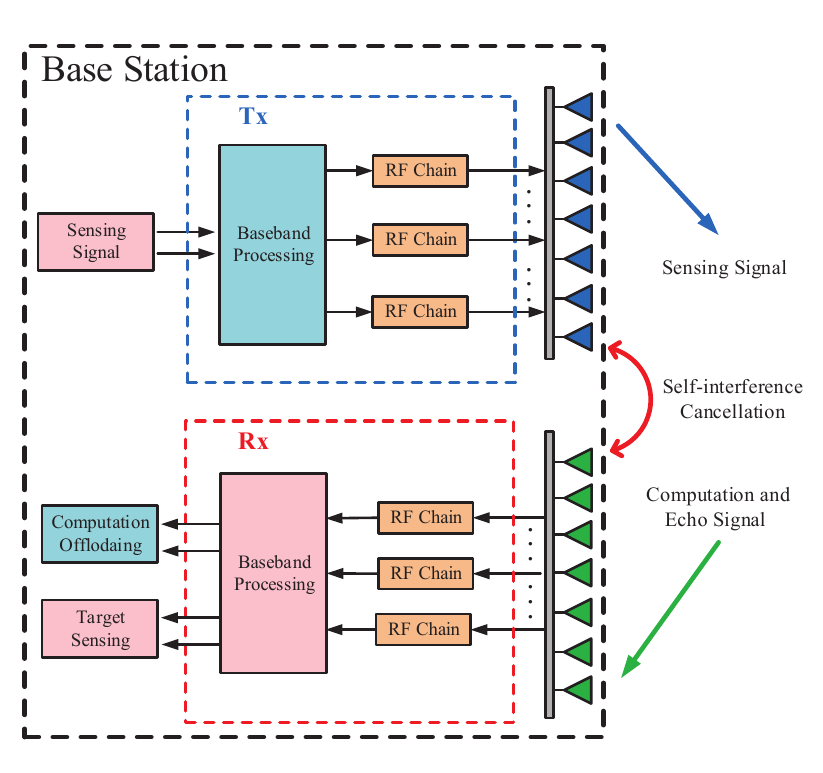}
\caption{The structure of the full duplex integrated sensing, computing, and communication base station.}
\label{FDBS}
\end{figure}

As shown in Fig.~\ref{system_model}, a STAR-RIS aided ISCC system for IoRT is proposed, where the full-duplex (FD) BS equipped with an edge server simultaneously provides edge computing services for $L$ decision robots (DRs) and senses a target robot (TR) in its vicinity. The DR refers to the robot equipped with various sensors such as temperature sensors, humidity sensors, magnetic field sensors, etc., and has the ability to autonomously make decisions. The TR refers to the robot that the BS intends to sense. {The block fading channel model is adopted, where the channel characteristics remain constant within each coherence block and vary between different coherence blocks. The robots are assumed to be stationary within each channel coherence time.} The FD BS is equipped with $N_t$ transmit antennas and $N_r$ receive antennas. All DRs are equipped with single antenna. The direct links between the BS and DRs are blocked with obstacles and an $N$-element STAR-RIS is deployed to establish extra communication links between them. The indices for DRs and RIS elements are collected in sets $\mathcal{L}=\{1,2,\cdots,L\}$ and $\mathcal{N}=\{1,2,\cdots,N\}$, respectively. The indices for DRs in the reflection region and the transmission region of the STAR-RIS are collected in sets $\mathcal{L}_r$ and $\mathcal{L}_t$, respectively. The FD BS sends the downlink sensing through the transmit uniform linear array (ULA) and collects the uplink offloading signals from DRs as well as the echo signal from the TR using a receive ULA. The structure of the considered FD BS is given in Fig.~\ref{FDBS}.

\subsection{STAR-RIS Model}
Unlike conventional RIS with non-magnetic elements, the advanced electric and magnetic elements in STAR-RIS can re-radiate the signals impinging on it in both reflection manner and transmission manner. For the $n$-th element in the RIS, define reflection coefficient $\phi_n^t=\sqrt{\rho_{n}^t}e^{j\theta_{n}^t}$ and transmission coefficient $\phi_n^r=\sqrt{\rho_{n}^r}e^{j\theta_{n}^r}$ to characterize its reflection and transmission performance. Specifically, $\rho_{n}^t$ and $\rho_{n}^r$ are the amplitude coefficients of the $n$-th STAR element, while $\theta_n^t\in [0,2\pi )$ and $\theta_n^r\in [0,2\pi )$ are the phase coefficients of the $n$-th STAR element. With coefficients defined above, the transmission and reflection matrices of the STAR-RIS can be given as $\mathbf{\Phi}_{i}=\diag\left(\sqrt{\rho_{1}^i}e^{j\theta_{1}^i}, \sqrt{\rho_{2}^i}e^{j\theta_{2}^i}, \cdots, \sqrt{\rho_{N}^i}e^{j\theta_{N}^i}\right), \forall i\in\{t,r\}$.
Since the STAR-RIS is a passive equipment, the amplitude coefficients of each element should satisfy the law of conservation of energy, which can be expressed as
\begin{equation}
    \rho _n^t + \rho _n^r = 1, \rho _n^i \ge 0, \forall i\in\{t,r\}.
\end{equation}
\subsection{Signal Model}
The signal transmitted from the BS is given as
\begin{equation}
    \mathbf{x}=\mathbf{w}d,
\end{equation}
where $\mathbf{w} \in \mathbb{C}^{N_t\times 1}$ represents the sensing beamformer. ${d}$ is the sensing signal with normalized power. {The power consumption for sensing at the BS can be given as $P_d = \mathbb{E}\{\mathbf{x}^H\mathbf{x}\}=\mathbf{w}^H\mathbf{w}$.}
% \begin{equation}
%     {P_d = \mathbb{E}\{\mathbf{x}^H\mathbf{x}\}=\mathbf{w}^H\mathbf{w}.}
% \end{equation}

The signal received at the BS contains the uplink offloading signals from the DRs, the echo sensing signal, and the self-interference signal. It can be expressed as
\begin{equation}
\begin{aligned}
    \tilde{\mathbf{y}}=\underbrace{\sum_{l=1}^{L}\sqrt{P_u}\mathbf{H}^H\mathbf{\Phi}(l)\mathbf{h}_{u,l}c_l}_{\text{uplink offloading signal}}&+ \underbrace{{\mathbf{A}}\mathbf{x}}_{\text{echo sensing signal}}\\
     &+ \underbrace{\mathbf{H}_{SI}\mathbf{x}}_{\text{self-interference}}+{\mathbf{n}},
\end{aligned}
\end{equation}
where $\mathbf{H}\in \mathbb{C}^{N\times N_r}$ is the channel from the BS to the RIS. $\mathbf{h}_{u,l}\in \mathbb{C}^{N\times 1}, \forall l \in \mathcal{L},$ is the channel from the DR $l$ to the RIS. $c_l\in \mathbb{C}, \forall l \in \mathcal{L},$ is the information symbol from DR $l$ with normalized power. $P_u$ is the transmit power of DRs. ${\mathbf{A}}$ and $\mathbf{H}_{SI}$ are the sensing channel and the self-interference channel between two ULAs at the BS, respectively. ${\mathbf{n}}\sim \mathcal{CN}(\mathbf{0},\sigma_u^2\mathbf{I})$ is the AWGN at the receiver of the BS and $\sigma_u^2$ is the noise power. $\mathbf{\Phi}(l)=\mathbf{\Phi}_r$ when $l\in \mathcal{L}_r$; $\mathbf{\Phi}(l)=\mathbf{\Phi}_t$ when $l\in \mathcal{L}_t$.

The difficulty for FD BS design lies in that the transmit and receive antenna arrays are co-located at the FD BS. The outgoing signal from the BS may overwhelm its receive hardware with limited-dynamic-range, making it challenging to extract useful signals at the BS~\cite{6280258}. Solutions like hardware isolation, analog and digital cancellation are proposed to carry out self-interference cancellation~\cite{6832464,8642523}, which can achieve $100$ dB SI suppression~\cite{9737357}. Thus, the self-interference after suppression is omitted in this work. The signal received at the BS after self-interference cancellation is given by
\begin{equation}
    \mathbf{y}=\sum_{l=1}^{L}\sqrt{P_u}\mathbf{H}^H\mathbf{\Phi}(l)\mathbf{h}_{u,l}c_l+{\mathbf{A}}\mathbf{x}+{\mathbf{n}}.
\end{equation}

Note that the TR is assumed to be located in the vicinity of the BS and the channels between the BS and the TR can be modeled with line-of-sight (LoS) channels. The antenna spacing of the ULAs adopted at the BS is half wavelength. When detecting the TR at an angle of interesting $\theta_0$, the array response vector between the BS transmitting antenna and the TR is~\cite{stutzman2012antenna}
\begin{equation}
    \mathbf{a}_t(\theta_0)=\frac{1}{\sqrt{N_t}}[1,e^{j\sin(\theta_0)},\cdots,e^{j(N_t-1)\sin(\theta_0)}]^T.
\end{equation}
The array response vector between the BS receiving antenna and the TR is
\begin{equation}
    \mathbf{a}_r(\theta_0)=\frac{1}{\sqrt{N_r}}[1,e^{j\sin(\theta_0)},\cdots,e^{j(N_r-1)\sin(\theta_0)}]^T.
\end{equation}
The sensing channel between the BS and the target can be expressed as
\begin{equation}
    {\mathbf{A}}_0=\alpha_0\mathbf{a}_r(\theta_0)\mathbf{a}_t(\theta_0)^H,
\end{equation}
where  $\alpha_0$ is the complex sensing coefficient decided by the radar cross-section of the TR and the path loss between the BS and the TR. Assume there are $M$ interferes distributed across $M$ different angles, i.e., $\theta_1,\theta_2,\cdots,\theta_M$~\cite{liu2022joint}. The sensing channel between the BS and interferes can be expressed as
\begin{equation}
    {\mathbf{A}}_I=\sum_m\mathbf{A}_m=\sum_m\alpha_m\mathbf{a}_r(\theta_m)\mathbf{a}_t(\theta_m)^H,
\end{equation}
where $\alpha_m$ is the complex sensing coefficient decided by the radar cross-section of the $m$-th interfere and the path loss between the BS and the $m$-th interfere.

The sensing channel can be expressed as
\begin{equation}
    {\mathbf{A}}={\mathbf{A}}_0+{\mathbf{A}}_I=\sum_{m=0}^{M}{\mathbf{A}}_m.
\end{equation}

 In this work, the RIS is deployed near the DRs which are blocked by obstacles to establish communication links for them. The TR is in the vicinity of the BS. Therefore, the RIS-related sensing link may not exist due to an unfavorable wireless path between the RIS and the TR. Besides, even if the RIS-related sensing link exists, the sensing signals need to experience the BS-RIS-targt-BS path to reach the BS. The long propagation distance and the double fading introduced by the RIS lead to limited echo signal power. As a result, the sensing signal reflected by the RIS is omitted at the receiver of the BS in this work.
\subsection{Communication Model and Sensing Model}
In this section, the communication model for the uplink computation offloading for DRs and the sensing model for probing the TR are introduced. Assign receive beamformer $\mathbf{u}_l$ for extracting information for DR $l$ from received signal $\mathbf{y}$ at the BS. The uplink communication SINR for DR $l$ can be expressed as in~\eqref{gamma_up} at the top of the next page, wherein
\begin{equation}
    \mathbf{R}_l=P_u\sum_{i\ne l}\mathbf{H}^H\mathbf{\Phi}(i)\mathbf{h}_{u,i}^H\mathbf{h}_{u,i}^H\mathbf{H}\mathbf{\Phi}(i)+\mathbf{A}\mathbf{w}\mathbf{w}^H\mathbf{A}^H+\sigma_u^2\mathbf{I}_{N_r}.
\end{equation}
\begin{figure*}[!t]
\normalsize
\begin{equation}\label{gamma_up}
\begin{aligned}
    \gamma_{u,l}&=\frac{P_u\mathbb{E}\{|\mathbf{u}_l^H\mathbf{H}^H\mathbf{\Phi}(l)\mathbf{h}_{u,l}c_l|^2\}}{P_u\sum_{i\ne l}\mathbb{E}\{|\mathbf{u}_l^H\mathbf{H}^H\mathbf{\Phi}(i)\mathbf{h}_{u,i}c_i|^2\}+\mathbb{E}\{|\mathbf{u}_l^H\mathbf{A}\mathbf{x}|^2\}+\mathbb{E}\{|\mathbf{u}_l^H\mathbf{n}|^2\}}\\
    &=\frac{P_u|\mathbf{u}_l^H\mathbf{H}^H\mathbf{\Phi}(l)\mathbf{h}_{u,l}|^2}{P_u\sum_{i\ne l}|\mathbf{u}_l^H\mathbf{H}^H\mathbf{\Phi}(i)\mathbf{h}_{u,i}|^2+\mathbf{u}_l^H\mathbf{A}\mathbf{w}\mathbf{w}^H\mathbf{A}^H\mathbf{u}_{l}+\sigma_u^2\|\mathbf{u}_l\|_2^2}\\
    &=\frac{P_u\mathbf{u}_l^H\mathbf{H}^H\mathbf{\Phi}(l)\mathbf{h}_{u,l}\mathbf{h}_{u,l}^H\mathbf{\Phi}(l)^H\mathbf{H}\mathbf{u}_l}{\mathbf{u}_l^H\mathbf{R}_l^H\mathbf{u}_{l}}.
\end{aligned}
\end{equation}
\hrulefill \vspace*{0pt}
\end{figure*}
The achievable communication rate for DR $l$ can be given as
\begin{equation}\label{R_l}
    R_{u,l} = \log(1+\gamma_{u,l}), \forall l \in \mathcal{L}.
\end{equation}

% \subsection{Sensing Model}
{In  this work, we adopt conventional BS sensing for the sensing function.} Upon extracting the uplink offloading signals from DRs, the BS carries out successive interference cancellations to remove the decoded offloading signals from the received signal~\cite{6182560}. Assuming perfect successive interference cancellation at the BS, the signal for target sensing can be given as
\begin{equation}
    \mathbf{y}_s={\mathbf{A}}\mathbf{x} +{\mathbf{n}}.
\end{equation}
{In sensing systems, the sensing accuracy of the target is often a monotonically increasing function of the sensing SINR~\cite{cui2013mimo}. Thus, we adopt the sensing SINR as the performance measure for the sensing feature, which can be expressed as
\begin{equation}\label{sensing_SINR}
    \gamma_{rad}=\frac{\mathbf{u}^H\mathbf{A}_0\mathbf{w}\mathbf{w}^H\mathbf{A}_0^H\mathbf{u}}{\mathbf{u}^H\mathbf{A}_I\mathbf{w}\mathbf{w}^H\mathbf{A}_I^H\mathbf{u}+\sigma_u^2\|\mathbf{u}\|_2^2},
\end{equation}
where $\mathbf{u}$ represents the receive beamformer for target sensing. The sensing performance of the system can be ensured by setting a relatively large sensing SINR requirement.}%\cite{9724174}The mobility of the decision robot affects the channel coherence time of off-loading channels.

% { The robots in IoRT systems are typically mobile. For communication and computation, we use the block fading channel model, where the channel remains constant within the coherence block and changes over coherent blocks. Our focus is on communicating and offloading within specific channel coherence time. When the speed of the decision robot increases, the corresponding channel coherence time shortens~\cite{cheng2008doppler}. For the sensing function, the Doppler effect caused by the movement can complicate the target sensing process. Our focus is on target sensing within a specific range-Doppler bin. However, the proposed technique can also be applied when the target's speed is high, as the suggested framework can be applied to each Doppler bin separately~\cite{1703855}.}

\subsection{Computation Model}
In this section, we introduce the computation model for the proposed ISCC system for the IoRT. Due to limited computation ability and power, the DRs offload their computation tasks to the edge server at the BS through uplink transmission links. {Compared to binary offloading, partial offloading has the benefit of making greater use of parallel computing resources and available bandwidth. It enables flexible component/data splitting, and by assigning time- or energy-consuming sub-tasks to MEC servers, it can result in increased energy savings and decreased computation delays~\cite{mao2017survey}.} The delay caused by the computing at the edge server and the return of the computation results to the DRs is omitted considering the superior computing power of the edge server and the limited size of the computation results. 

The edge server allocates CPU-cycle frequency (cycle/s) $f_l$ for the DR $l$ to execute a computation task that requires $\phi$ computation cycles to process one bit of data. Under this configuration, the computation rate of DR $l$ at the BS can be given by
\begin{equation}
    r_l = \frac{f_l}{\phi}, \forall l \in \mathcal{L}.
\end{equation}
{The power required to execute the computation task for DR $l$ can be given as}
\begin{equation}
    {P_l = \kappa f_l^3 = \kappa({\phi}r_l)^3, \forall l \in \mathcal{L},}
\end{equation}
where $\kappa$ is the power coefficient decided by the CPU chip architecture at the edge server. Note that the communication capacity between the DRs and the BS is limited, the computation rate of DRs should satisfy the following constraints
\begin{equation}\label{comm_comp}
   r_l \le BR_{u,l}, \forall l \in \mathcal{L},
\end{equation}
where $B$ is the communication bandwidth for uplink offloading. Constraint~\eqref{comm_comp} arises from the communication-computation causality principle, whereby the maximum number of bits processed by the edge server for DR $l$ must not surpass the transmission capacity of the link connecting the BS and DR $l$. The sum computation rate of DRs is 
\begin{equation}
    R = \sum_{l=1}^{L}r_l.
\end{equation}

\section{FD ISCC Design for Computation Rate Maximization} 
\subsection{Optimization Problem Formulation}
In this work, we aim to maximize the sum computation rate of DRs by optimizing the sensing beamformer $\mathbf{w}$, receive beamformers $\mathbf{u}_l,\forall l,$ and the STAR-RIS reflection matrices $\mathbf{\Phi}_r$, $\mathbf{\Phi}_t$. The sum computation maximization problem can be given as
\begin{subequations}\label{problem:sum_comp_rate}
    \begin{align}        
        \max_{\mathcal{A}} \quad &  R \\
        \label{constraint:SINR_rad}
        \mathrm{s.t.} \quad  & \gamma_{rad} \ge \Gamma_{rad},\\
        \label{constraint:off}
        & r_l \le BR_{u,l}, \forall l \in \mathcal{L},\\
        \label{constraint:power}
        &P_d+\sum_{l=1}^{L}P_l\le P_b ,\\
        \label{constraint:RIS1}
        & \theta_n^i \in (0,2\pi], \forall n\in \mathcal{N}, i \in\{t,r\},\\
        \label{constraint:RIS}
        & \rho _n^i \in [0,1], \forall n\in \mathcal{N}, i \in\{t,r\},\\
        \label{constraint:RIS3}
        & \rho _n^t + \rho _n^r = 1, \forall n\in \mathcal{N},
    \end{align}
\end{subequations}
where $\mathcal{A}=\{\mathbf{w},\mathbf{u}, \{\mathbf{u}_l\}_{l=1}^{L}, \{r_l\}_{l=1}^{L},\mathbf{\Phi}_r,\mathbf{\Phi}_t\}$ is the set for optimization variables. {$P_b$ is the power budget of the BS. $P_d$ and $P_l$ are the power consumption
for sensing and the power consumption
for executing the computation task for DR $l$, respectively.} Constraint~\eqref{constraint:SINR_rad} ensures that the minimum SINR for radar operations exceeds the designated sensing threshold $\Gamma_{rad}$. Constraint~\eqref{constraint:off} ensures that the computation bits processed by the edge server for DR $l$ are less than the communication capacity of the offloading link. Constraint~\eqref{constraint:power} limits the power consumption for sensing and computation under the power budget of the BS. Constraints~\eqref{constraint:RIS1}-\eqref{constraint:RIS3} are the STAR-RIS coefficients constraint. 
\begin{remark} \textbf{offloading-sensing trade-off:} \emph{Both target sensing and offloading computation consume the limited power budget at the BS, giving rise to an inherent offloading-sensing trade-off.  As more data is offloaded to the BS's edge server, the edge server consumes more power. Consequently, less power is available for target sensing, which can lead to unsatisfying sensing performance. Conversely, allocating more power to target sensing reduces the power available for computing the offloaded data at the BS, which may result in degraded offloading performance.}
\end{remark}

\subsection{Equivalent Communication Model for the Sensing Constraint}
% to reformulate the sensing constraint~\eqref{constraint:SINR_rad}.
% An equivalent multiple-input and multiple-output (MIMO) communication system is proposed where the SINR of the proposed MIMO system has the same form as the sensing SINR of the ISCC system. 

{An equivalent multiple-input and multiple-output (MIMO) communication system is proposed, wherein the SINR of the MIMO system has the same form as the sensing SINR of the ISCC system. Then, we replace the sensing SINR constraint~\eqref{constraint:SINR_rad} with the equivalent MIMO communication capacity constraint.} Assuming there is a $N_t$-antenna BS communicates with a $N_r$-antenna user, the signal received by the user can be given as
\begin{equation}
     \mathbf{y}_{rad}=\mathbf{A}_0\mathbf{w}d+\mathbf{n}_{rad},
 \end{equation}
 where $d$ is the transmitted signal of the equivalent communication system and $\mathbf{w}$ is the beamformer for the transmitted signal. $\mathbf{A}_0$ is the channel between the BS and the user. $\mathbf{n}_{rad}$ is the noise vector with covariance matrix $\mathbf{R}_{rad}=\mathbf{A}_I\mathbf{w}\mathbf{w}^H\mathbf{A}_I+\sigma_u^2\mathbf{I}_{N_r}$. The capacity of the equivalent communication system can be expressed as~\cite{tse2005fundamentals}
 \begin{equation}
\begin{aligned}\label{R_rad}
    R_{rad}=&\log\left(1+\frac{\mathbf{u}^H\mathbf{A}_0\mathbf{w}\mathbf{w}^H\mathbf{A}_0^H\mathbf{u}}{\mathbf{u}^H\mathbf{A}_I\mathbf{w}\mathbf{w}^H\mathbf{A}_I^H\mathbf{u}+\sigma_u^2\|\mathbf{u}\|_2^2}\right)\\
     =&\log(1+\gamma_{rad}).
\end{aligned}
\end{equation}
Define $r_{rad}=\log(1+\Gamma_{rad})$. The constraint~\eqref{constraint:SINR_rad} can be reformulated as 
\begin{equation}
    r_{rad}\le R_{rad}.
\end{equation}
The sum computation maximization problem can be given as
\begin{subequations}\label{problem:sum_comp_rate0}
    \begin{align}        
        \max_{\mathcal{A}} \quad &  R \\
        \label{constraint:SINR_rad0}
        \mathrm{s.t.} \quad  & r_{rad}\le R_{rad},\\
        & \eqref{constraint:off}-\eqref{constraint:RIS3},
    \end{align}
\end{subequations}
{where constraints \eqref{constraint:off}-\eqref{constraint:RIS3} are the DR offloading constraint, the power constraint of the BS, and the STAR-RIS coefficients constraint.}

The AO method is adopted to tackle problem~\eqref{problem:sum_comp_rate0}, where the variables $\mathcal{B}=\{\mathbf{w},\mathbf{u}, \{\mathbf{u}_l\}_{l=1}^{L}, \{r_l\}_{l=1}^{L}\}$ is optimized with fixed variables $\mathcal{C}=\{\mathbf{\Phi}_r,\mathbf{\Phi}_t\}$, and vice versa.

\subsection{Optimization Problem with respect to $\mathcal{B}$}

With fixed STAR-RIS matrix coefficients $\mathbf{\Phi}_r$ and $\mathbf{\Phi}_t$, the sum computation rate maximization problem can be expressed as
\begin{subequations}\label{problem:sum_comp_rate1}    
    \begin{align}    
        \max_{\mathcal{B}} \quad &  \sum_{l=1}^{L}r_l \\
        \mathrm{s.t.} \quad  & \eqref{constraint:SINR_rad0},\eqref{constraint:off},\\
        \label{constraint:power1}
        &\mathbf{w}^H\mathbf{w}+\sum_{l=1}^{L}\kappa({\phi}r_l)^3\le P_b.
    \end{align}
\end{subequations}
Problem~\eqref{problem:sum_comp_rate1} is non-convex since constraints \eqref{constraint:SINR_rad0} and \eqref{constraint:off}  are non-convex. To tackle the non-convex constraint \eqref{constraint:off}, the WMMSE method is adopted~\cite{5756489}. With $\mathbf{u}$ and $\mathbf{u}_l$ decoding the signal of the equivalent communication system and the offloading signals from the DR $l$, respectively, the decoded signals for the equivalent communication system and the DR $l$ can be expressed as
\begin{equation}
    \hat{d}=\mathbf{u}^H\mathbf{y}_{rad},
\end{equation}
\begin{equation}
    \hat{c}_l=\mathbf{u}_l^H\mathbf{y},
\end{equation}
The mean square error (MSE) of the signal of the equivalent communication system and the signal transmitted from DR $l$ are
\begin{equation}\label{e_rad}
    e_{rad}=\mathbb{E}[|\hat{d}-{d}|^2]=\mathbf{u}^H\mathbf{R}_1\mathbf{u}-2\Re(\mathbf{u}^H\mathbf{A}_0\mathbf{w})+1,
\end{equation}
\begin{equation}\label{e_l}
    e_l=\mathbb{E}[|\hat{c}_l-{c}_l|^2]=\mathbf{u}_l^H\mathbf{R}_2\mathbf{u}_l-2\sqrt{P_u}\Re(\mathbf{u}_l^H\mathbf{H}^H\mathbf{\Phi}(l)\mathbf{h}_{u,l})+1,
\end{equation}
where $\mathbf{R}_1$ is the covariance matrix of the signal received at the equivalent communication system, which can be expressed as

\begin{equation}
\begin{aligned}
    \mathbf{R}_1=&\mathbb{E}[\mathbf{y}_{rad}\mathbf{y}_{rad}^H]\\
    =&\mathbf{A}_0\mathbf{w}\mathbf{w}^H\mathbf{A}_0^H+\mathbf{A}_I\mathbf{w}\mathbf{w}^H\mathbf{A}_I^H+\sigma_u^2\mathbf{I}_{N_r}.
\end{aligned}
\end{equation}
$\mathbf{R}_2$ is the covariance matrix of the signal received at the BS of the ISCC system, which can be expressed as
\begin{equation}
\begin{aligned}
    \mathbf{R}_2=&\mathbb{E}[\mathbf{y}\mathbf{y}^H]\\
    =&\sum_{l}\mathbf{H}^H\mathbf{\Phi}(l)\mathbf{h}_{u,l}\mathbf{h}_{u,l}^H\mathbf{\Phi}(l)^H\mathbf{H}+\mathbf{A}\mathbf{w}\mathbf{w}^H\mathbf{A}^H+\sigma_u^2\mathbf{I}_{N_r}.
\end{aligned}
\end{equation}
% Define $\eta_l=\log(\lambda_l)-\lambda e_{l}+1$ with $\lambda_l \in \mathbb{R}^+, \forall l\in \mathcal{L}$. and $\eta_l$ 

In the following proposition, the tight lower bounds of the capacities of the equivalent communication system as well as the uplink offloading are formulated using the rate-MMSE relationship.  and the capacity of the uplink offloading of DR~$l$, respectively.
\begin{proposition}
\emph{Define $\eta=\log(\lambda)-\lambda e_{rad}+1$ with auxiliary variable $\lambda \in \mathbb{R}^+$. $\eta$ servers as a tight lower bound for the capacity of the equivalent communication system. $\eta$ satisfies:}
\begin{equation}\label{optimization:eta}
    \max_{\lambda,\mathbf{u}}\eta = R_{rad},
\end{equation}
\begin{proof}
Maximizing $\eta$ with respect to $\mathbf{u}$ is equivalent to minimizing $e_{rad}$, i.e., the MSE of the equivalent communication system. The MMSE receive beamformer $\mathbf{u}$ can be obtain by setting $\partial e_{rad}/\partial {\mathbf{u}} = 0$, which lead to
\begin{equation}\label{u_opt}
    \mathbf{u}^*=\mathbf{R}_1^{-1}\mathbf{A}_0\mathbf{w}
\end{equation}
Substituting~\eqref{u_opt} into~\eqref{e_rad}, the MMSE of the equivalent communication system can be expressed as
\begin{equation}\label{e_rad_MMSE}
\begin{aligned}
    e_{rad}^{\text{MMSE}}=&1-\mathbf{w}^H\mathbf{A}_0^H\mathbf{R}_1^{-1}\mathbf{A}_0\mathbf{w}\\
    =&(1+\mathbf{w}^H\mathbf{A}_0^H\mathbf{R}_{rad}^{-1}\mathbf{A}_0\mathbf{w})^{-1}.
\end{aligned}
\end{equation}
Substituting~\eqref{u_opt} into~\eqref{R_rad}, the capacity of the equivalent communication system can be expressed as
\begin{equation}
\begin{aligned}
    R_{rad}=\log(1+\mathbf{w}^H\mathbf{A}_0^H\mathbf{R}_{rad}^{-1}\mathbf{A}_0\mathbf{w}).
\end{aligned}
\end{equation}
With given MMSE receive beamformer $\mathbf{u}^{\text{MMSE}}$, the $\lambda$ maximizing $\eta$ can be obtained by setting $\partial \eta/\partial {\lambda} = 0$, which lead to
\begin{equation}\label{lambda_opt}
     \lambda^*=(e_{rad}^{\text{MMSE}})^{-1}
 \end{equation} 
and 
\begin{equation}\label{eta_opt}
    \max_{\lambda,\mathbf{u}}\eta =\log((e_{rad}^{\text{MMSE}})^{-1}).
\end{equation}
Substituting~\eqref{e_rad_MMSE} into~\eqref{eta_opt}, the equation~\eqref{optimization:eta} is established.
\end{proof}
\end{proposition}

\begin{proposition}
\emph{Define $\eta_l=\log(\lambda_l)-\lambda e_{l}+1$ with auxiliary variable $\lambda_l \in \mathbb{R}^+, \forall l\in \mathcal{L}$. $\eta_l$ servers as a tight lower bound for the capacity of the uplink offloading of DR $l$. $\eta_l$ satisfies:}
\begin{equation}\label{optimization:eta_l}
    \max_{\lambda_l,\mathbf{u}_l}\eta_l = R_{u,l}, \forall l,
\end{equation}
\begin{proof}
The proof of proposition 2 is similar to the proof of proposition 1 and is omitted. Besides, the expression of the optimal receive beamformer and auxiliary variable for DR $l$ is given as follows
\begin{equation}\label{u_l_opt}
    \mathbf{u}_l^*=\mathbf{R}_2^{-1}\mathbf{H}^H\mathbf{\Phi}(l)\mathbf{h}_{u,l}, \forall l,
\end{equation}
\begin{equation}\label{e_l_MMSE}
\begin{aligned}
    e_{rad}^{\text{MMSE}}=&1-\mathbf{h}_{u,l}^H\mathbf{\Phi}(l)\mathbf{H}\mathbf{R}_2^{-1}\mathbf{H}^H\mathbf{\Phi}(l)\mathbf{h}_{u,l}\\
    =&(1+\mathbf{h}_{u,l}^H\mathbf{\Phi}(l)\mathbf{H}\mathbf{R}_{l}^{-1}\mathbf{H}^H\mathbf{\Phi}(l)\mathbf{h}_{u,l})^{-1}, \forall l,
\end{aligned}
\end{equation}
\begin{equation}\label{lambda_l_opt}
    \lambda_l^*=(e_{l}^{\text{MMSE}})^{-1}, \forall l,   
\end{equation}
\end{proof}
\end{proposition}

With~\eqref{optimization:eta} and~\eqref{optimization:eta_l}, problem~\eqref{problem:sum_comp_rate1} can be reformulated as
\begin{subequations}\label{problem:sum_comp_rate2}
    \begin{align}        
        \max_{\hat{\mathcal{B}}} \quad &  \sum_{l=1}^{L}r_l \\
        \mathrm{s.t.} \quad  & r_{rad}\le \eta,\\
        & r_{l}\le \eta_l, \forall l,\\
        &\eqref{constraint:power1},
    \end{align}
\end{subequations}%$\mathcal{O}\left(I_1\max\{L+1,N_t\}^4N_t^{\frac{1}{2}}\right)$
where $\hat{\mathcal{B}}=\{\mathbf{w},\mathbf{u}, \lambda, \{\mathbf{u}_l\}_{l=1}^{L}, \{\lambda_l\}_{l=1}^{L}, \{r_l\}_{l=1}^{L}\}$. When the receive beamformers and auxiliary variables given by~\eqref{u_opt},~\eqref{lambda_opt},~\eqref{u_l_opt}, and~\eqref{lambda_l_opt}, problem~\eqref{problem:sum_comp_rate2} is convex with respect to $\{\mathbf{w}, \{r_l\}_{l=1}^{L}\}$, which can be solved with CVX~\cite{grant2014cvx}. {The proposed algorithm for solving problem~\eqref{problem:sum_comp_rate2} is summarized in~\textbf{Algorithm~\ref{algorithm1}}. The computational complexity of calculating MMSE beamformers and auxiliary variables is $\mathcal{O}\left(2(L+1)N_r^3\right)$~\cite{horn2012matrix}. The computational complexity of solving convex quadratically constrained problem in step $4$ is $\mathcal{O}\left((L+1)^{0.5}\left((L+1)N_t^2+N_t^3\right)\right)$~\cite{nesterov1994interior}. The computational complexity of~\textbf{Algorithm~\ref{algorithm1}} is given by $\mathcal{O}\left(I_1\left(2(L+1)N_r^3+(L+1)^{0.5}\left((L+1)N_t^2+N_t^3\right)\right)\right)$,~where $I_1$ is the iteration number for~\textbf{Algorithm~\ref{algorithm1}} to converge.} The convergence of~\textbf{Algorithm~\ref{algorithm1}} is guaranteed by the following proposition.
% \begin{algorithm}[t]
% \caption{The Proposed Algorithm for Solving Problem~\eqref{problem:sum_comp_rate2}.}\label{algorithm1}
% \begin{algorithmic}[1]
% \STATE {Set initial sensing beamformer $\mathbf{w}^{(0)}$, $i=0$.}\\
% \STATE {\bf repeat: }\\
% \STATE \quad Calculate $\mathbf{u}^{(i+1)}$ and $\mathbf{u}_l^{(i+1)}, \forall l,$ based on~\eqref{u_opt} and~\eqref{u_l_opt}.\\
% \STATE \quad Calculate $\lambda^{(i+1)}$ and $\lambda_l^{(i+1)}, \forall l,$ based on~\eqref{lambda_opt} and~\eqref{lambda_l_opt}.\\
% \STATE \quad Solve problem~\eqref{problem:sum_comp_rate2} with obtained receive beamformers and auxiliary variables.\\
% \STATE \quad Update $\{\mathbf{w}^{(i+1)}, \{r_l^{(i+1)}\}_{l=1}^{L}\}$.\\
% \STATE \quad Set $i =i+1$.\\
% \STATE {\bf until} {the fractional increase of the sum computation
% rate is less than the convergence threshold}.
% \end{algorithmic}
% \end{algorithm}
\begin{algorithm}[t]
\caption{{The Proposed Algorithm for Solving Problem~\eqref{problem:sum_comp_rate2}.}}\label{algorithm1}
\begin{algorithmic}[1]
{\STATE {Set initial sensing beamformer $\mathbf{w}^{(0)}$, $i=0$.}\\
\STATE  Calculate $\mathbf{u}^{(i+1)}$ and $\mathbf{u}_l^{(i+1)}, \forall l,$ based on~\eqref{u_opt} and~\eqref{u_l_opt}.\\
\STATE  Calculate $\lambda^{(i+1)}$ and $\lambda_l^{(i+1)}, \forall l,$ based on~\eqref{lambda_opt} and~\eqref{lambda_l_opt}.\\
\STATE  Solve problem~\eqref{problem:sum_comp_rate2} with obtained receive beamformers and auxiliary variables.\\
\STATE  Update $\{\mathbf{w}^{(i+1)}, \{r_l^{(i+1)}\}_{l=1}^{L}\}$.\\
\STATE {If the fractional increase of the objective function is less than the convergence threshold, terminate. Otherwise, set $i =i+1$ and go to step 2}.}
\end{algorithmic}
\end{algorithm}
\begin{proposition}
\emph{{Algorithm~\ref{algorithm1}} is guaranteed to converge to a stationary point of problem~\eqref{problem:sum_comp_rate2}.}
\begin{proof}
The sum computation rate of the ISCC system is bounded above due to the limited power budget at the BS. During the $i$-th iteration of the \textbf{Algorithm~\ref{algorithm1}}, the sum computation rate satisfies~\eqref{convergence} at the top of the next page. It can be observed that the sum computation rate is non-
decreasing over every iteration. As a result, the proposed algorithm can reach at least a stationary point of problem~\eqref{problem:sum_comp_rate2}.
\begin{figure*}[!t]
\normalsize
\begin{equation}\label{convergence}
    \begin{aligned}
        R&\left(\mathbf{w}^{(i)}, \{r_l^{(i)}\}_{l=1}^{L}, \mathbf{u}^{(i)}, \lambda^{(i)}, \{\mathbf{u}^{(i)}_l\}_{l=1}^{L}, \{\lambda^{(i)}_l\}_{l=1}^{L}\right)\\ 
        &= R\left(\mathbf{w}^{(i)}, \{r_l^{(i)}\}_{l=1}^{L}, \mathbf{u}^{(i+1)}, \lambda^{(i+1)}, \{\mathbf{u}^{(i+1}_l\}_{l=1}^{L}, \{\lambda^{(i+1)}_l\}_{l=1}^{L}\right)\\
        &\le R\left(\mathbf{w}^{(i+1)}, \{r_l^{(i+1)}\}_{l=1}^{L}, \mathbf{u}^{(i+1)}, \lambda^{(i+1)}, \{\mathbf{u}^{(i+1)}_l\}_{l=1}^{L}, \{\lambda^{(i+1)}_l\}_{l=1}^{L}\right),
    \end{aligned}
\end{equation}
\hrulefill \vspace*{0pt}
\end{figure*}
\end{proof}
\end{proposition}

\subsection{Optimization Problem with respect to $\mathcal{C}$}
With given variables $\mathcal{B}=\{\mathbf{w},\mathbf{u}, \{\mathbf{u}_l\}_{l=1}^{L}, \{r_l\}_{l=1}^{L}\}$, the sum computation rate maximization problem can be expressed as
\begin{subequations}\label{problem:sum_comp_rate3}
    \begin{align}        
        \text{Find} \quad &  {\mathcal{C}} \\
        \label{constraint:SINR_off3}
        \mathrm{s.t.} \quad  & \gamma_{u,l}\ge \Gamma_{u,l}, \forall l,\\
        & \eqref{constraint:RIS1}-\eqref{constraint:RIS3},
    \end{align}
\end{subequations}
where $\Gamma_{u,l}=e^{\frac{r_l}{B}}-1$.
Define
\begin{equation}
\begin{aligned}
\mathbf{v}(l)=&\text{diag}(\mathbf{\Phi}(l)),\\
=&\left[\sqrt{\rho_1^i}e^{j\theta_1^i}, \sqrt{\rho_2^i}e^{j\theta_2^i}, \cdots, \sqrt{\rho_N^i}e^{j\theta_N^i}\right]^T, \forall i \in\{t,r\},
\end{aligned}
\end{equation}
which leads to $\mathbf{v}(l)=\text{diag}(\mathbf{\Phi}_r)$ when $l\in \mathcal{L}_r$; $\mathbf{\Phi}(l)=\text{diag}(\mathbf{\Phi}_t)$ when $l\in \mathcal{L}_t$. Constraint~\eqref{constraint:SINR_off3} can be reformulated as
\begin{equation}
\begin{aligned}
    &\frac{P_u}{\Gamma_{u,l}}\mathbf{u}_l^H\mathbf{H}^H\text{diag}(\mathbf{h}_{u,l})\mathbf{v}(l)\mathbf{v}(l)^H\text{diag}(\mathbf{h}^H_{u,l})\mathbf{H}\mathbf{u}_l\ge\\
    &\sum_{i\ne l}{P_u}\mathbf{u}_l^H\mathbf{H}^H\text{diag}(\mathbf{h}_{u,i})\mathbf{v}(i)\mathbf{v}(i)^H\text{diag}(\mathbf{h}^H_{u,i})\mathbf{H}\mathbf{u}_l+N_l, \forall l, 
    \end{aligned}
\end{equation}
where $N_l=\mathbf{u}_l^H\mathbf{A}\mathbf{w}\mathbf{w}^H\mathbf{A}^H\mathbf{u}_{l}+\sigma_u^2\|\mathbf{u}_l\|_2^2$.
Define $\mathbf{V}(l)=\mathbf{v}(l)\mathbf{v}(l)^H, \forall l$. Constraint~\eqref{constraint:SINR_off3} can be further reformulated as
\begin{equation}\label{re_SINR_off3}
\begin{aligned}
    \frac{1}{\Gamma_{u,l}}\text{Tr}\left(\mathbf{B}_l^l\mathbf{V}(l)\right)\ge \sum_{i\ne l}\text{Tr}\left(\mathbf{B}_l^i\mathbf{V}(i)\right)+N_l, \forall l, 
\end{aligned}
\end{equation}
where $\mathbf{B}_l^i=P_u\text{diag}(\mathbf{h}^H_{u,i})\mathbf{H}\mathbf{u}_l\mathbf{u}_l^H\mathbf{H}^H\text{diag}(\mathbf{h}_{u,i}), \forall l,i\in \mathcal{L}$.

\begin{subequations}\label{problem:sum_comp_rate4}
    \begin{align}        
        \text{Find} \quad &  \mathbf{V}_i,\forall i \in \{t,r\} \\
        \mathrm{s.t.} \quad  & \eqref{re_SINR_off3},\\
        \label{rank}
        & \text{rank}(\mathbf{V}_i) = 1, \forall i \in \{t,r\},\\
        \label{RIS1}
        & \mathbf{V}_i \succeq 0, \forall i \in \{t,r\},\\
        \label{RIS2}
        & \text{diag}(\mathbf{V}_t+\mathbf{V}_r)=\mathbf{I}_N.
    \end{align}
\end{subequations}
The problem~\eqref{problem:sum_comp_rate4} is non-convex due to the non-convex constraint~\eqref{rank}, which is equivalent to the following format 
\begin{equation}\label{norm}
    \|\mathbf{V}_i\|_*-\|\mathbf{V}_i\|_2= 0, \forall i,
\end{equation}
where $\|\mathbf{V}_i\|_*=\sum_j\sigma_j(\mathbf{V}_i)$ with $\sigma_j(\mathbf{V}_i)$ representing the $j$-th singular value (in descending order) of $\mathbf{V}_i$. $\|\mathbf{V}_i\|_2=\sigma_1(\mathbf{V}_i)$ is the spectral norm of $\mathbf{V}_i$. For any semi-definite Hermitian matrix $\mathbf{V}_i$, constraint~\eqref{norm} is satisfied only when the constraint~\eqref{rank} is met. The penalty method is used to tackle the constraint~\eqref{norm}~\cite{doi:10.1137/S1052623493259215,mu2021simultaneously}. Introducing penalty coefficients $\rho_i, \forall i \in \{t,r\},$ the problem~\eqref{problem:sum_comp_rate4} can be reformulated as
\begin{subequations}\label{problem:sum_comp_rate5}
    \begin{align}        
        \min_{\mathbf{V}_r,\mathbf{V}_t} \quad &   \sum\nolimits_{i\in\{t,r\}}\rho_i(\|\mathbf{V}_i\|_*-\|\mathbf{V}_i\|_2)\\
        \mathrm{s.t.} \quad  & \eqref{re_SINR_off3}, \eqref{RIS1}, \eqref{RIS2}.
    \end{align}
\end{subequations}
\begin{algorithm}[t]
\caption{{ The Proposed Algorithm for Solving Problem~\eqref{problem:sum_comp_rate5}.}}\label{algorithm2}
\begin{algorithmic}[1]
{ \STATE {Set initial STAR-RIS coefficients $\mathbf{V}_r^{(0)}$ and  $\mathbf{V}_t^{(0)}$, $n=0$.}\\
\STATE  Solve problem~\eqref{problem:sum_comp_rate6} with feasible $\mathbf{V}_r^{(n)}$ and  $\mathbf{V}_t^{(n)}$.\\
\STATE  Update $\mathbf{V}_r^{(n+1)}$ and $\mathbf{V}_t^{(n+1)}$.\\
\STATE {If the fractional decrease of the objective function is less than the convergence threshold, terminate. Otherwise, set $n =n+1$ and go to step 2}.}
\end{algorithmic}
\end{algorithm}
According to~\cite[Theorem 1]{9183907}, when $\rho_i$ is sufficiently large, the solution to problem~\eqref{problem:sum_comp_rate5} is also feasible to problem~\eqref{problem:sum_comp_rate4}. Thus, in the following, the problem~\eqref{problem:sum_comp_rate5} with large penalty coefficients are solved instead of problem~\eqref{problem:sum_comp_rate4}.  Problem~\eqref{problem:sum_comp_rate5} is still non-convex due to its non-convex object function. By adopting successive convex approximation (SCA) with any feasible $\mathbf{V}_r^{(n)},\mathbf{V}_t^{(n)}$, the non-convex terms in the object function of problem~\eqref{problem:sum_comp_rate5} can be approximated as
\begin{equation}
\begin{aligned}
    -\|\mathbf{V}_i\|_2 &\le -\|\mathbf{V}_i^{(n)}\|_2-\text{Tr}\left(\mathbf{b}_i^{(n)}(\mathbf{b}_i^{(n)})^H(\mathbf{V}_i-\mathbf{V}_i^{(n)})\right)\\
    &={V}_i^{(n)},
\end{aligned}
\end{equation}
where $\mathbf{b}_i^{(n)}$ is the eigenvector of $\mathbf{V}_i^{(n)}$ corresponding to the singular value $\sigma_1(\mathbf{V}_i^{(n)})$. Substituting ${V}_i^{(n)}, \forall i,$ in to the objective function, in the $n$-th iteration of SCA, the surrogate problem of problem~\eqref{problem:sum_comp_rate5} can be expressed as
\begin{subequations}\label{problem:sum_comp_rate6}
    \begin{align}        
        \min_{\mathbf{V}_r,\mathbf{V}_t} \quad &   \sum_{i\in\{t,r\}}\rho_i(\|\mathbf{V}_i\|_*+{V}_i^{(n)})\\
        \mathrm{s.t.} \quad  & \eqref{re_SINR_off3}, \eqref{RIS1}, \eqref{RIS2}.
    \end{align}
\end{subequations}
% \begin{algorithm}[t]
% \caption{The Proposed Algorithm for Solving Problem~\eqref{problem:sum_comp_rate5}.}\label{algorithm2}
% \begin{algorithmic}[1]
% \STATE {Set initial STAR-RIS coefficients $\mathbf{V}_r^{(0)}$ and  $\mathbf{V}_t^{(0)}$, $n=0$.}\\
% \STATE {\bf repeat: }\\
% \STATE \quad Solve problem~\eqref{problem:sum_comp_rate6} with feasible $\mathbf{V}_r^{(n)}$ and  $\mathbf{V}_t^{(n)}$.\\
% \STATE \quad Update $\mathbf{V}_r^{(n+1)}$ and $\mathbf{V}_t^{(n+1)}$.\\
% \STATE \quad Set $n =n+1$.\\
% \STATE {\bf until} {the fractional increase of the objective function is less than the convergence threshold}.
% \end{algorithmic}
% \end{algorithm}
Problem~\eqref{problem:sum_comp_rate6} is convex with respect to $\mathbf{V}_r$ and $\mathbf{V}_t$ and can be solved with CVX. {The proposed algorithm for solving problem~\eqref{problem:sum_comp_rate5} is summarized in~\textbf{Algorithm~\ref{algorithm2}}. The computational complexity of~\textbf{Algorithm~\ref{algorithm2}} mainly comes from solving semidefinite programming (SDP) in step $2$, which has the computational complexity of $\mathcal{O}\left(I_2\max\{L+2,N\}^4N^{\frac{1}{2}}\right)$~\cite{5447068}. $I_2$ is the iteration number for~\textbf{Algorithm~\ref{algorithm2}} to converge.} The convergence of~\textbf{Algorithm~\ref{algorithm2}} is guaranteed by the following proposition.
\begin{proposition}
\emph{{Algorithm~\ref{algorithm2}} is guaranteed to converge to a stationary point of problem~\eqref{problem:sum_comp_rate5}.}
\begin{proof}
Note that the objective function of~\eqref{problem:sum_comp_rate5} is bounded above by the objective function of~\eqref{problem:sum_comp_rate6}. Each iteration of~\textbf{Algorithm~\ref{algorithm2}} gives a non-increasing objective function of~\eqref{problem:sum_comp_rate6}. Thus, the proposed~\textbf{Algorithm~\ref{algorithm2}} can at least converge to a stationary point of~\eqref{problem:sum_comp_rate5}.
\end{proof}
\end{proposition}
\begin{algorithm}[t]
\caption{{ The Proposed Algorithm for Solving Problem~\eqref{problem:sum_comp_rate0}.}}\label{algorithm3}
\begin{algorithmic}[1]
{\STATE {Set initial $\mathbf{w}^{(0)}$, $\mathbf{V}_r^{(0)}$ and  $\mathbf{V}_t^{(0)}$, $j=0$.}\\
\STATE  Solve problem~\eqref{problem:sum_comp_rate2} with fixed $\mathcal{C}^{(j)}$, initial sensing beamformer with $\mathbf{w}^{(j)}$.\\
\STATE  Update $\mathcal{B}^{(j+1)}$.\\
\STATE  Solve problem~\eqref{problem:sum_comp_rate4} with fixed $\mathcal{B}^{(j+1)}$, initial STAR-RIS coefficients with $\mathbf{V}_r^{(j)}$ and  $\mathbf{V}_t^{(j)}$.\\
\STATE  Update $\mathcal{C}^{(j+1)}$.\\
\STATE {If the fractional increase of the objective function is less than the convergence threshold, terminate. Otherwise, set $j =j+1$ and go to step 2}.}
\end{algorithmic}
\end{algorithm}
\subsection{The Overall Algorithm}
{The proposed algorithm for solving problem~\eqref{problem:sum_comp_rate0} is summarized in~\textbf{Algorithm~\ref{algorithm3}}, where $\mathcal{B}^{(j)}=\big\{\mathbf{w}_0^{(j)}, \mathbf{u}^{(j)}, \{\mathbf{u}_l^{(j)}\}_{l=1}^{L}, \{r_l^{(j)}\}_{l=1}^{L}\big\}$ and $\mathcal{C}^{(j)}=\big\{\mathbf{V}_r^{(j)}, \mathbf{V}_t^{(j)}\big\}$. The computational complexity of~\textbf{Algorithm~\ref{algorithm3}} is given by $\mathcal{O}(I_3(I_1\left(2(L+1)N_r^3+(L+1)^{0.5}\left((L+1)N_t^2+N_t^3\right)\right)+I_2\max\{L+2,N\}^4N^{\frac{1}{2}}))$, where $I_3$ is the iteration number for~\textbf{Algorithm~\ref{algorithm3}} to converge. Thanks to the generality of the STAR-RIS, the sum computation rate maximization problem for traditional RIS aided ISCC systems can also be solved with the proposed algorithms, resulting in comparable computation complexity.}

\begin{proposition}
\emph{{Algorithm~\ref{algorithm3}} is guaranteed to converge to a stationary point of problem~\eqref{problem:sum_comp_rate0}.}
\begin{proof}
 At the $(j-1)$-th iteration of the \textbf{Algorithm~\ref{algorithm3}}, the step $4$ gives objective function $R\left(\mathcal{B}^{(j)},\mathcal{C}^{(j)}\right)$. Based on~\textbf{Proposition~3}, updating $\mathcal{B}$ with fixed $\mathcal{C}$ can at least give a non-decreasing objective function. It follows that $R\left(\mathcal{B}^{(j)},\mathcal{C}^{(j)}\right)\le R\left(\mathcal{B}^{(j+1)},\mathcal{C}^{(j)}\right)$. 

 At the $j$-th iteration of the \textbf{Algorithm~\ref{algorithm3}}, the step $2$ gives objective function $R\left(\mathcal{B}^{(j+1)},\mathcal{C}^{(j)}\right)$ and the step $4$ gives objective function $R\left(\mathcal{B}^{(j+1)},\mathcal{C}^{(j+1)}\right)$. It follows that $R\left(\mathcal{B}^{(j+1)},\mathcal{C}^{(j)}\right)=R\left(\mathcal{B}^{(j+1)},\mathcal{C}^{(j+1)}\right)$ since the objective function is regardless of STAR-RIS coefficients.

Then it follows that the sum computation rate is non-decreasing over every iteration of Algorithm 3, i.e.,
\begin{equation}
    R\left(\mathcal{B}^{(j)},\mathcal{C}^{(j)}\right) \le R\left(\mathcal{B}^{(j+1)},\mathcal{C}^{(j+1)}\right).
\end{equation}
Recall that the sum computation rate of the ISCC system is bounded above due to the limited power budget at the BS. The proposed algorithm can reach at least a stationary
point of problem~\eqref{problem:sum_comp_rate0}.
\end{proof}
\end{proposition}
{The proposed algorithm is carried out at the BS (equipped with the edge server), and the required information exchanges between the BS, RIS, and decision robots in the proposed ISCC system include}\footnote{{This work focuses on the uplink scenario, though the information exchanges between the BS and decision robots involve downlink transmission.}}{: $(1)$ The BS sends the coefficient configuration to the STAR-RIS via the dedicated controller connecting the BS and RIS. $(2)$ The BS feeds back the allocated offloading computation rate to each decision robot via downlink transmission links. $(3)$ The decision robots offload their computation tasks to the edge server at the BS through uplink transmission links. $(4)$ The BS returns the computation results to the decision robots via downlink transmission links.}
% \begin{algorithm}[t]
% \caption{The Proposed Algorithm for Solving Problem~\eqref{problem:sum_comp_rate0}.}\label{algorithm3}
% \begin{algorithmic}[1]
% \STATE {Set initial $\mathbf{w}^{(0)}$, $\mathbf{V}_r^{(0)}$ and  $\mathbf{V}_t^{(0)}$, $j=0$.}\\
% \STATE {\bf repeat: }\\
% \STATE \quad Solve problem~\eqref{problem:sum_comp_rate2} with fixed $\mathcal{C}^{(j)}$, initial sensing beamformer $\mathbf{w}^{(j)}$.\\
% \STATE \quad Update $\mathcal{B}^{(j+1)}$.\\
% \STATE \quad Solve problem~\eqref{problem:sum_comp_rate4} with fixed $\mathcal{B}^{(j+1)}$, initial STAR-RIS coefficients $\mathbf{V}_r^{(j)}$ and  $\mathbf{V}_t^{(j)}$.\\
% \STATE \quad Update $\mathcal{C}^{(j+1)}$.\\
% \STATE \quad Set $j =j+1$.\\
% \STATE {\bf until} {the fractional increase of the objective function is less than the convergence threshold}.
% \end{algorithmic}
% \end{algorithm}
\section{Simulation Results} \label{sec:results}
In this section, the simulation setup of the proposed ISCC IoRT system is introduced. Then, simulation results are provided to verify the effectiveness of the proposed algorithm for solving the sum computation rate maximization problem. 
\subsection{Simulation Setup}
\begin{figure}[!htbp]
\centering
\includegraphics[width=0.5\textwidth]{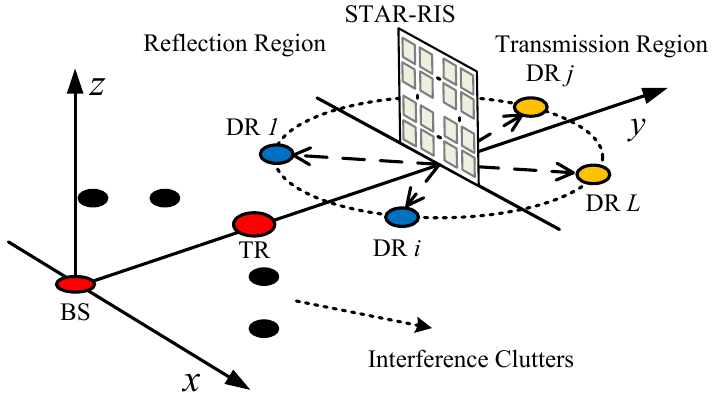}
\caption{Simulation setup for the STAR-RIS aided ISCC IoRT system.}
\label{setup}
\end{figure}
Fig.~\ref{setup} depicts the simulation setup for the considered STAR-RIS aided ISCC IoRT system, where the BS is placed at the original point of the coordinate system and the STAR-RIS is placed at $(0,20)$ (m). DRs are spread around the STAR-RIS and fall on the circles centered at the STAR-RIS with a radius of $5$ m. The BS is assumed to be equipped with $N_t=8$ transmit antennas and $N_r=8$ receive antennas. The number of STAR-RIS elements is set as $N=40$. The number of DRs $L$ is set as an even number and all DRs are equally divided into the transmission and reflection sides of the STAR-RIS. The TR is set at $(0,10)$ (m), while $M=4$ interfere clusters fall on the circles centered at the original point with a radius of $10$ m. The angles of interfere clusters are $\{\theta_1, \theta_2, \theta_3, \theta_4\}=\{-\frac{\pi}{3}, -\frac{\pi}{6}, \frac{\pi}{6}, \frac{\pi}{3}\}$.

The STAR-RIS can be deployed at a favorable position where the LoS channels can be established between both the BS and the STAR-RIS as well as the STAR-RIS and DRs~\cite{hua2022throughput}. Thus, the channels between the BS and the STAR-RIS as well as the channels between the STAR-RIS and DRs can be modeled with Rician fading, which can be expressed as
as follows:
\begin{equation}\label{Rician}
\mathbf{h}=\sqrt{\beta}\left(\sqrt{\frac{\epsilon}{\epsilon+1}}\mathbf{h}_\text{LoS}+\sqrt{\frac{1}{\epsilon+1}}\mathbf{h}_\text{NLoS}\right),
\end{equation}
where $\mathbf{h}_\text{LoS}$ is the LoS channel component and the $\mathbf{h}_\text{NLoS}$ is the NLoS channel component. $\epsilon$ is the Rician factor. $\beta=Dd^{-\alpha}$ represents the path loss coefficient, where $D=-30$ dB is the path loss at the reference distance $1$ m. $d$ is the transmission distance and $\alpha$ is the path loss exponent. When it comes to channels between the BS and the TR or interfere clusters, the Rician channel model is similar to~\eqref{Rician} with different path loss coefficient $\beta_s$ which can be expressed as $\beta_s=D(2d)^{-\alpha}$.

The penalty coefficients adopted in problem~\eqref{problem:sum_comp_rate5} is $\rho_r=\rho_t=10^3$~\cite{9183907}. The communication bandwidth for uplink offloading is set as $B=20$ MHz. The power coefficient of the edge server is set as $\kappa=10^{-26}$~\cite{9996408}. The computation cycles required to process one bit of data at the edge server are set as $\phi=3\times 10^3$ cycle/bit. The noise power at the BS is $-90$ dBm. The transmit power of DRs is $P_u=10$ dBm. The sensing threshold is $\Gamma_\mathrm{rad}=30$ dB. The power budget at the BS is $P_b=30$ dBm. The Rician factor is set as $\epsilon=3$ dB and the path loss factor is set as $\alpha = 2.2$.
\subsection{Baseline Schemes}
To verify the effectiveness and benefits of the proposed STAR-RIS aided ISCC IoRT system, we consider the following baseline schemes.
\begin{enumerate}

\item \textbf{Conventional RIS Scheme:} In this baseline scheme, we adopt two conventional RISs (one reflection RIS and one transmission RIS) deployed adjacent to each other to realize full space coverage. To provide a fair comparison, each RIS contains $N/2$ elements. This benchmark can achieved with a STAR-RIS by setting $\mathbf{\rho}_n^t =1, \forall n\le N/2$, and $\mathbf{\rho}_n^r =1, \forall n > N/2$. The optimization problem that results from this configuration for RIS coefficients optimization can be given by

\begin{subequations}\label{problem:sum_comp_rate_s1}
    \begin{align}        
        \text{Find} \quad &  \mathbf{V}_i,\forall i \in \{t,r\} \\
        \mathrm{s.t.} \quad  & \eqref{re_SINR_off3}, \eqref{rank}, \eqref{RIS1},\\
        & \text{diag}(\mathbf{V}_t)=\text{Bdiag}\left\{\mathbf{I}_{\frac{N}{2}},\mathbf{0}_{\frac{N}{2}}\right\},\\
        & \text{diag}(\mathbf{V}_r)=\text{Bdiag}\left\{\mathbf{0}_{\frac{N}{2}},\mathbf{I}_{\frac{N}{2}}\right\}.
    \end{align}
\end{subequations}
This problem can be solved by Algorithm 2 with appropriate modification.

\item \textbf{Equal Splitting STAR-RIS Scheme:} In this baseline scheme, all STAR-RIS elements equally split the signal energy into the transmission and reflection sides. This benchmark can achieved with a STAR-RIS by setting $\mathbf{\rho}_n^t=\mathbf{\rho}_n^r =1/2, \forall n$. The optimization problem that results from this configuration for STAR-RIS coefficients optimization can be given by
\begin{subequations}\label{problem:sum_comp_rate_s2}
    \begin{align}        
        \text{Find} \quad &  \mathbf{V}_i,\forall i \in \{t,r\} \\
        \mathrm{s.t.} \quad  & \eqref{re_SINR_off3}, \eqref{rank}, \eqref{RIS1},\\
        & \text{diag}(\mathbf{V}_i)=\frac{1}{2}\mathbf{I}_N, \forall i \in \{t,r\}.
    \end{align}
\end{subequations}
This problem can be solved by Algorithm 2 with appropriate modification.

\item \textbf{Offloading Only Scheme:} In this baseline scheme, the BS only provides offloading computation service and does not carry out target sensing. Without the sensing to communication interference, the uplink communication SINR for DR $l$ can be expressed as
\begin{equation}
    \gamma_{u,l}=\frac{P_u|\mathbf{u}_l^H\mathbf{H}^H\mathbf{\Phi}(l)\mathbf{h}_{u,l}|^2}{P_u\sum_{i\ne l}|\mathbf{u}_l^H\mathbf{H}^H\mathbf{\Phi}(i)\mathbf{h}_{u,i}|^2+\sigma_u^2\|\mathbf{u}_l\|_2^2}.
\end{equation}
The optimization problem that results from this configuration can be given by

\begin{subequations}\label{problem:sum_comp_rate_s3}
    \begin{align}        
        \max_{\hat{\mathcal{A}}} \quad &  R \\
        \mathrm{s.t.} \quad  & \sum_{l=1}^{L}\kappa({\phi}r_l)^3\le P_b ,\\
        & \eqref{constraint:off}, \eqref{constraint:RIS1}, \eqref{constraint:RIS1}, \eqref{constraint:RIS1}, 
    \end{align}
\end{subequations}
where $\hat{\mathcal{A}}=\{\mathbf{u}, \{\mathbf{u}_l\}_{l=1}^{L}, \{r_l\}_{l=1}^{L},\mathbf{\Phi}_r,\mathbf{\Phi}_t\}$. This problem can be solved by Algorithm 3 with appropriate modification.

\end{enumerate}

In simulation figures, legends ``Proposed STAR-RIS'' represents the proposed STAR-RIS aided ISCC IoRT system. Legends  ``E-STAR-RIS'' represents ISCC IoRT system with {equal splitting} STAR-RIS. Legends ``C-RIS'' represents the ISCC IoRT system with conventional RIS. Legends ``Offloading Only'' represents the IoRT system without the sensing requirement.
\subsection{Simulation Results}
% \subsection{Convergence Behavior of the Proposed Algorithm}

\begin{figure}[!htbp]
\centering
\includegraphics[width=0.5\textwidth]{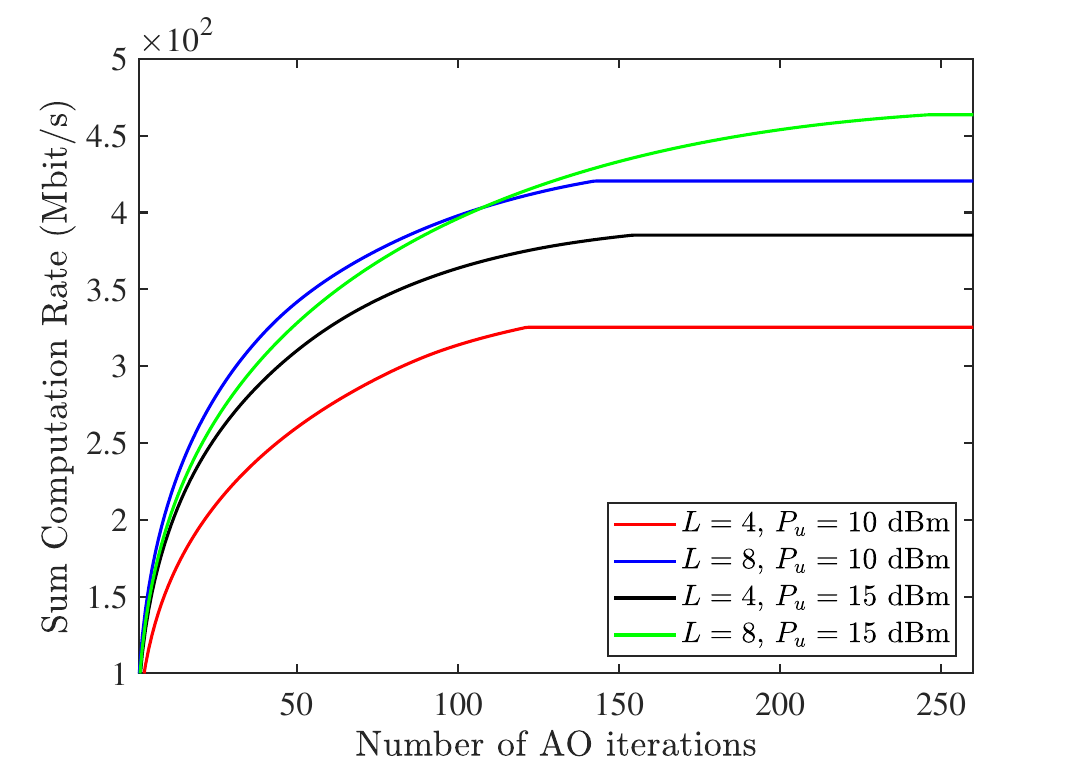}
\caption{The sum computation rate of DRs versus the number of AO iterations.}
\label{fig:con}
\end{figure}

The convergence of Algorithm 3 is analyzed in Fig.~\ref{fig:con}, illustrating the sum computation rate against the number of AO iterations. The number of DRs and their transmit power are configured as $L = 4$ or $8$ and $P_u = 10$ dBm or $15$~dBm, respectively. It is clear that the algorithms converge after several iterations, with the sum computation rates of all instances consistently increasing with each iteration.

\begin{figure}[!tbp]
\centering
\includegraphics[width=0.5\textwidth]{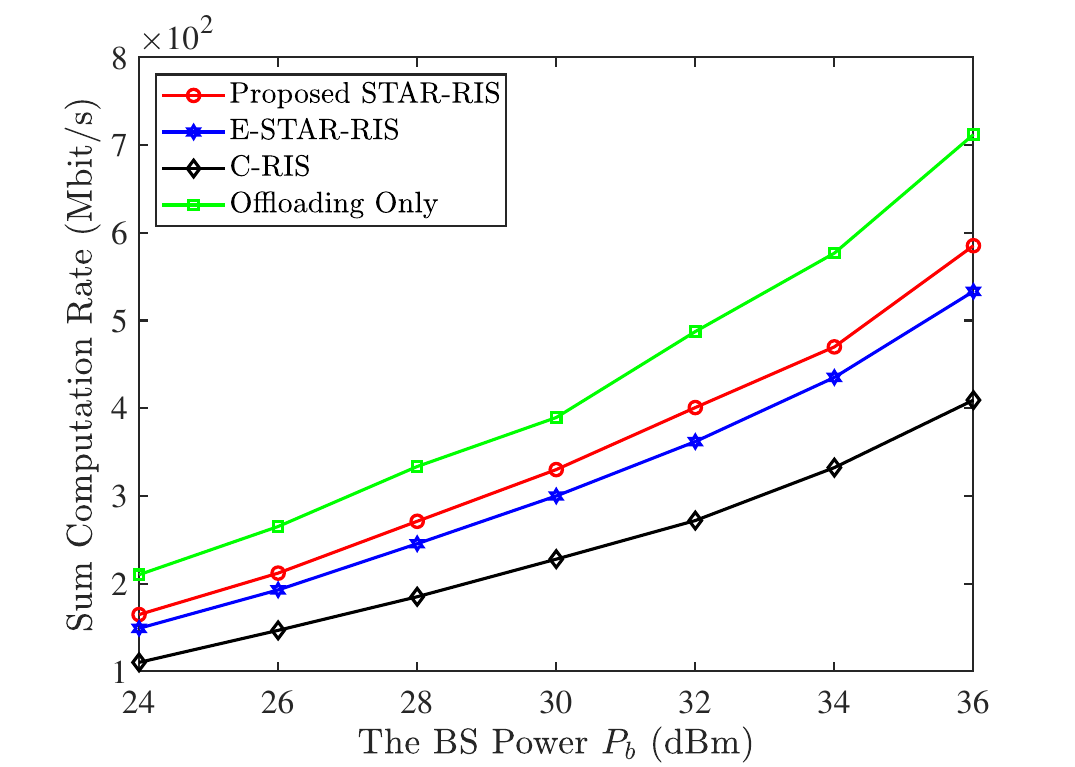}
\caption{The sum computation rate of DRs versus the power budget of the BS.}
\label{fig:P}
\end{figure}

In Fig.~\ref{fig:P}, the computation rates achieved by different schemes are examined in relation to the power budget at the BS, with a fixed sensing threshold of $\Gamma_s = 30$ dB. It is observed that the sum computation rate of all schemes increases as the power budget at the BS increases. This is due to the increased available power at the BS, enabling the edge server to process more data when computation is constrained by limited power. However, as the power at the BS continues to increase, the sum computation rate reaches a plateau. This plateau is attributed to the capacity limit of uplink offloading. The proposed schemes demonstrate superior performance compared to all ISCC schemes. This is expected because STAR-RIS effectively manages inter-user interference for DRs' uplink offloading and mitigates interference between sensing and communication, thereby enhancing computation performance within the ISCC system. The offloading-only scheme outperforms the ISCC schemes because all available power is allocated solely to the computation task at the BS.

\begin{figure}[!tbp]
\centering
\includegraphics[width=0.5\textwidth]{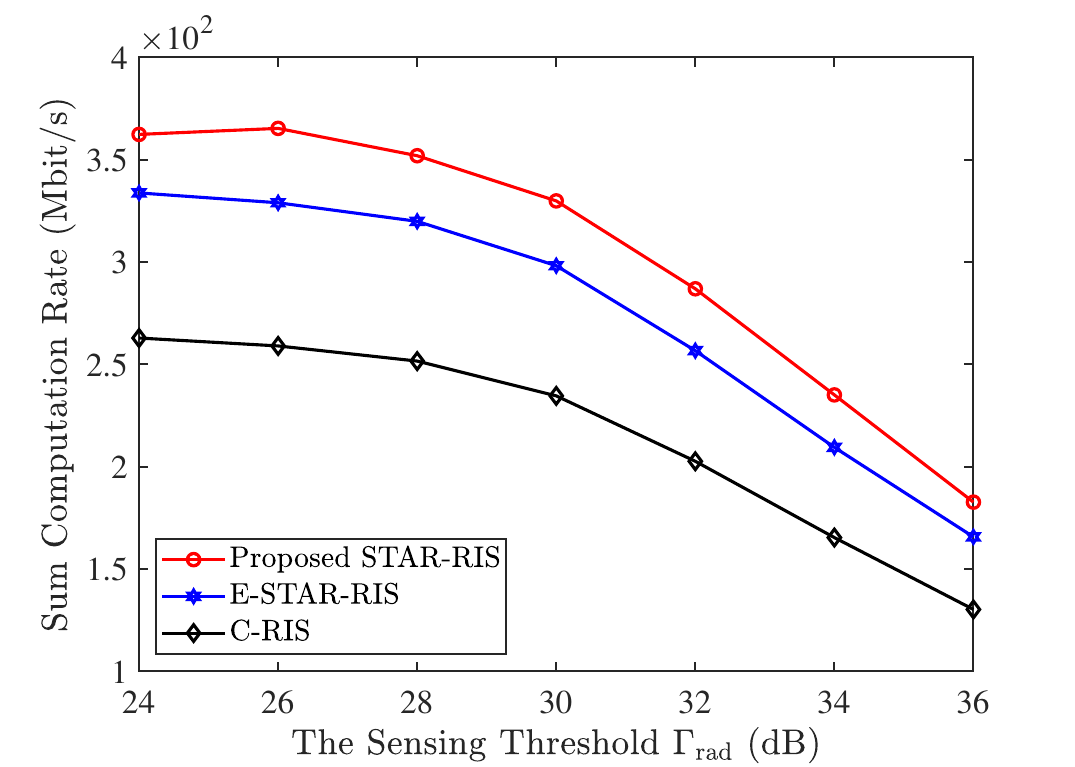}
\caption{The sum computation rate of DRs versus the sensing threshold.}
\label{fig:Gamma}
\end{figure}

Fig.~\ref{fig:Gamma} illustrates the sum computation rate plotted against the sensing threshold. The computation performance declines as the sensing threshold increases for all ISCC schemes. This trend is anticipated because allocating more power to the sensing function as the threshold increases constrains the available power for computation at the BS, reflecting the inherent offloading-sensing trade-off.

\begin{figure}[!tbp]
\centering
\includegraphics[width=0.5\textwidth]{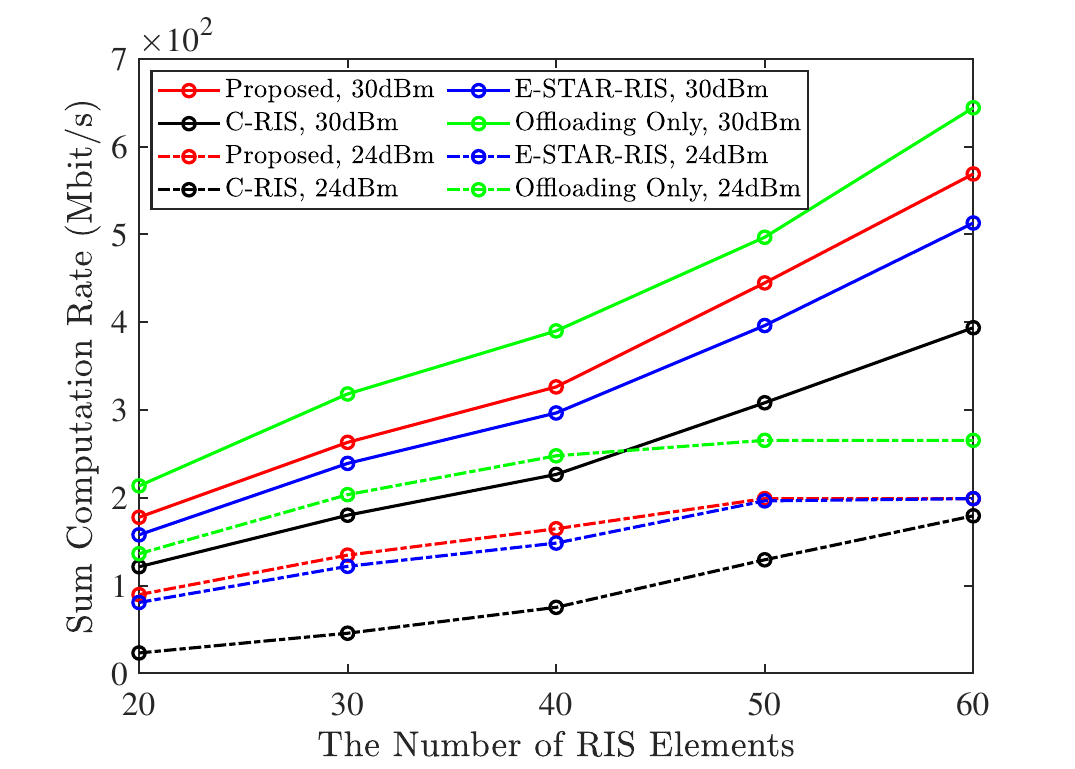}
\caption{The sum computation rate of DRs versus the number of RIS elements.}
\label{fig:N}
\end{figure}

 {Fig.~\ref{fig:N} depicts the total computation rate of DRs versus the number of RIS elements, where the power budget at the BS is set at $24$ dBm and $30$ dBm, respectively. It has been observed that as the number of RIS elements increases, the total computation rate also increases. Furthermore, an increase in the number of RIS elements leads to a larger performance difference between STAR-RIS and conventional RIS. This phenomenon occurs because more RIS elements provide additional opportunities to design more effective configuration strategies for STAR-RIS, thereby achieving greater performance benefits. The increase in computation rate brought by the addition of STAR-RIS elements is limited by the available power at the BS. When maintaining the minimum sensing SINR for the TR and performing computation at the edge server exhausts the BS's power budget, the computation rate at the BS reaches a plateau. This occurs because the BS lacks the capacity to provide additional computing resources, even though the increasing number of STAR-RIS elements can further enhance the offloading capacity between the BS and DRs.}

{In conclusion, when the computation rate at the BS is limited by the channel quality between the BS and DRs, increasing the number of STAR-RIS elements can effectively improve the computation rate. However, when the computation rate is limited by the power budget at the BS, further increasing the number of STAR-RIS elements does not provide additional benefits.}

\begin{figure}[!tbp]
\centering
\includegraphics[width=0.5\textwidth]{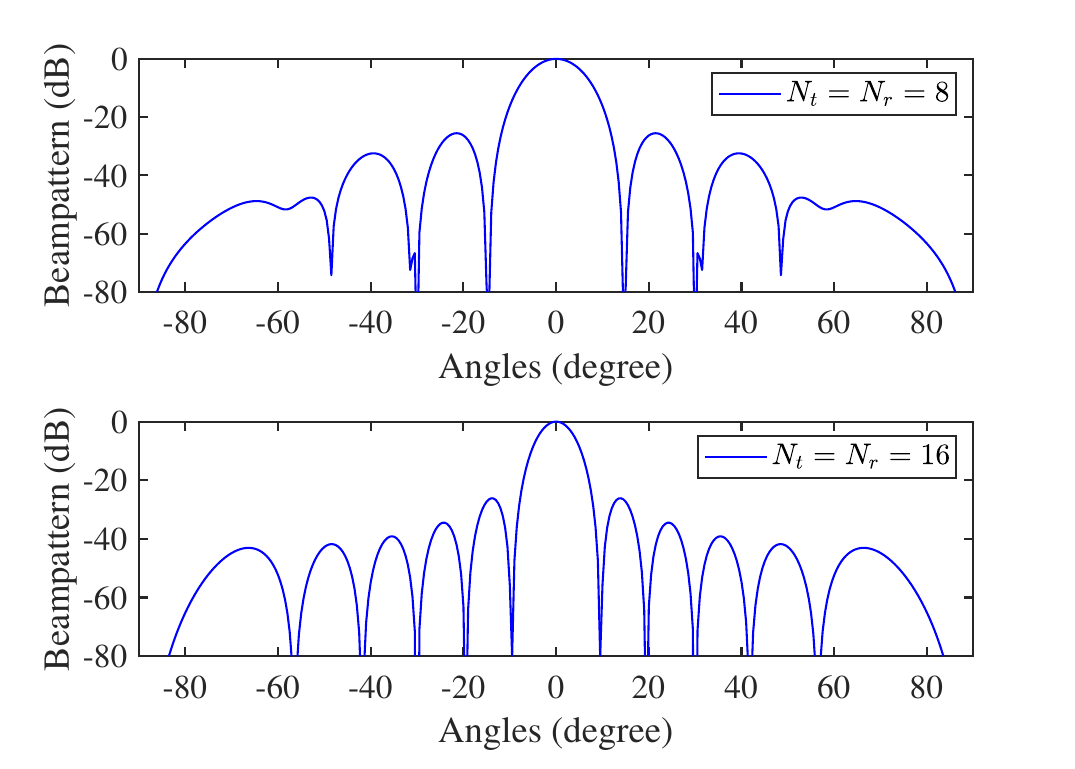}
\caption{The sensing beampattern achieved by Algorithm 3.}
\label{fig:B}
\end{figure}
{In Fig.~\ref{fig:B}, the sensing beampattern achieved by the proposed scheme with Algorithm 3. The normalized sensing beampattern at angle $\theta$ is defined as
\begin{equation}
    G(\theta)=\frac{\left|\mathbf{u}^H\mathbf{a}_r(\theta)\mathbf{a}_t(\theta)^H\mathbf{w}\right|^2}{\max_{\hat{\theta}}\left|\mathbf{u}^H\mathbf{a}_r(\hat{\theta})\mathbf{a}_t(\hat{\theta})^H\mathbf{w}\right|^2}.
\end{equation}
The main beam of the sensing beam pattern is focused on the TR, with several nulls deliberately positioned towards the interferers. This demonstrates the effectiveness of the proposed beamforming design for TR sensing  in the presence of interfering clusters. Moreover, the $16$-antenna system is capable of forming deeper nulls at the angles of the interferers.}
\section{Conclusion} 
This work proposed a STAR-RIS-aided ISCC IoRT framework, where the BS simultaneously senses the TR and computes data for DRs. To address the sum computation rate optimization problem, an AO algorithm was developed for beamforming design at the BS and coefficient optimization at the STAR-RIS. Simulation results were presented to demonstrate the convergence of the proposed algorithm. Additionally, the advantage of adopting STAR-RIS in the ISCC IoRT system was verified. {This work confirms the effectiveness of utilizing STAR-RIS to support communication and computation functions in ISCC IoRT systems. The potential of involving STAR-RIS not only in communication and computation but also in the sensing function can be an intriguing future research topic. Furthermore, given that the FD BS is adopted in the proposed ISCC systems, research on ISCC systems with simultaneous uplink offloading and downlink communication can be a promising direction for future exploration.}
% This work proposed a STAR-RIS aided ISCC IoRT framework, where the BS simultaneously senses the TR and computes data for DRs. To address the sum computation rate optimization problem, an AO algorithm was developed for beamforming design at the BS and coefficient optimization at the STAR-RIS. Simulation results were presented to demonstrate the convergence of the proposed algorithm. Besides, the advantage of adopting STAR-RIS in ISCC IoRT system was verified.

%  the sum computation rate depends on multiple parameters in the ISCC IoRT system, including the sensing threshold, the power available at the BS, and the number of RIS elements. Additionally, the advantage of adopting STAR-RIS in ISCC IoRT system is verified.
\bibliography{myref}
\bibliographystyle{IEEEtran}

\begin{IEEEbiography}[{\includegraphics[width=1in, height=1.25in,clip, keepaspectratio]{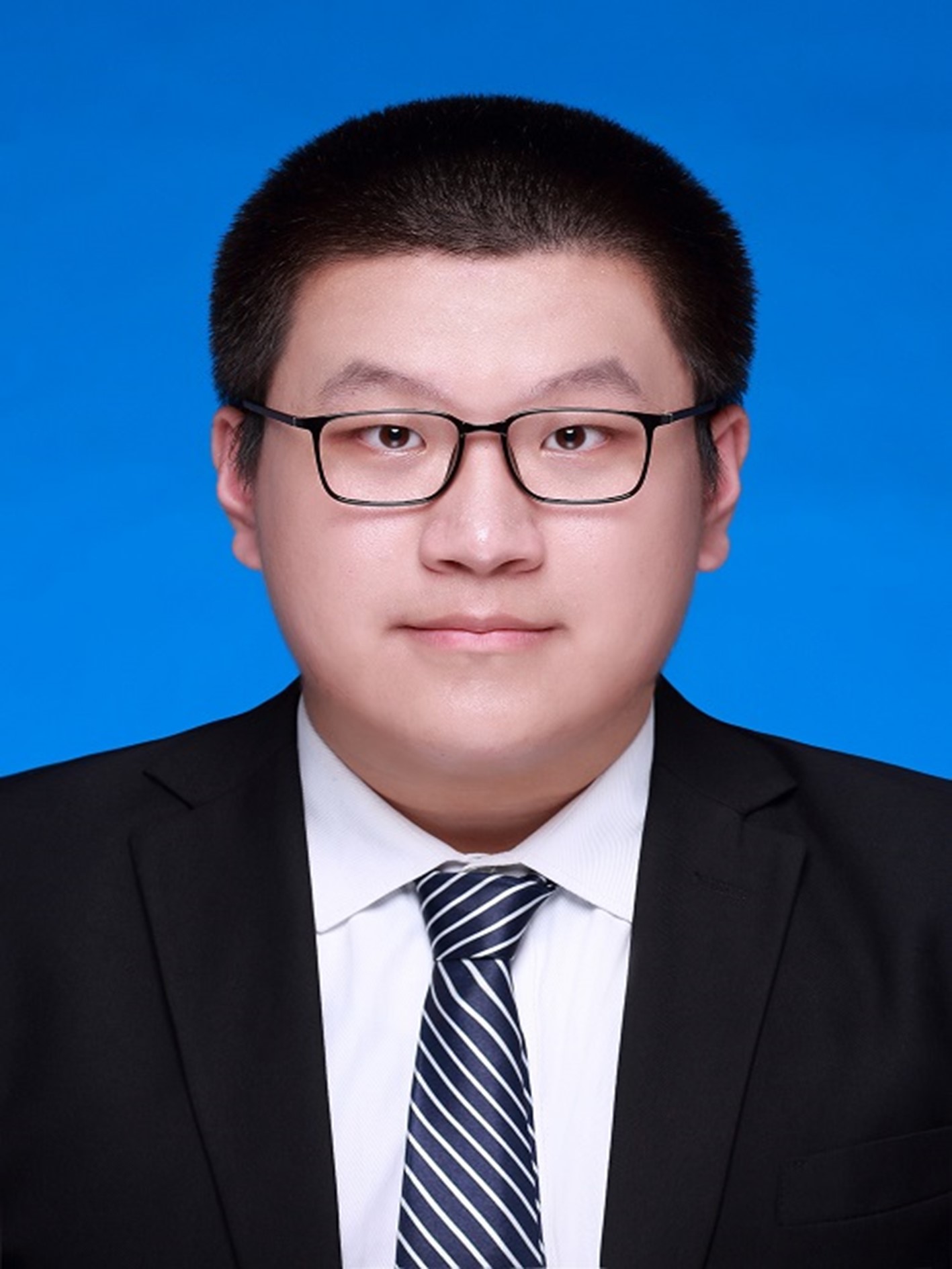}}]{Haochen Li} (Graduate Student Member, IEEE) received the B.S. degree in communication engineering from Nanjing University of Science and Technology, Nanjing, China, in 2019. He is currently pursuing the Ph.D. degree with the National Mobile Communications Research Laboratory, Southeast University, Nanjing, China. He is also a Research Associate with the School of Electronic Engineering and Computer Science, Queen Mary University of London, since September 2022. His current research interests include massive MIMO communication, reconfigurable intelligent surface, and integrated sensing and communications.
\end{IEEEbiography}

\begin{IEEEbiography}[{\includegraphics[width=1in, height=1.25in,clip, keepaspectratio]{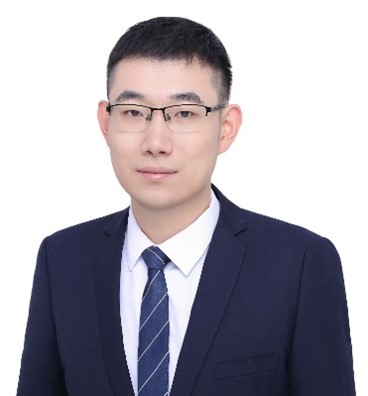}}]{Xidong Mu} (Member, IEEE, \url{https://xidongmu.github.io/}) received the Ph.D. degree in Information and Communication Engineering from the Beijing University of Posts and Telecommunications, Beijing, China, in 2022. 

He was with the School of Electronic Engineering and Computer Science, Queen Mary University of London, from 2022 to 2024, where he was a Postdoctoral Researcher. He has been a lecturer (an assistant professor) with the Centre for Wireless Innovation (CWI), Queen’s University Belfast, U.K. since August 2024. His research interests include non-orthogonal multiple access (NOMA), IRSs/RISs aided communications, integrated sensing and communications, semantic communications, and optimization theory. He received the IEEE ComSoc Outstanding Young Researcher Award for EMEA region in 2023, the Exemplary Reviewer Certificate of the \textsc{IEEE Transactions on Communications} in 2020 and 2022, and the \textsc{IEEE Communication Letters} in 2021-2023. He is the recipient of the 2024 IEEE Communications Society Heinrich Hertz Award, the Best Paper Award in ISWCS 2022, the 2022 IEEE SPCC-TC Best Paper Award, and the Best Student Paper Award in IEEE VTC2022-Fall. He serves as the Public Engagement and Social Networks Coordinator of IEEE ComSoC SPCC Technical Committee, the secretary of the IEEE ComSoc Next Generation Multiple Access (NGMA) Emerging Technology Initiative and the Special Interest Group (SIG) in SPCC Technical Committee on Signal Processing Techniques for NGMA. He also serves as an Editor of \textsc{IEEE Transactions on Communications}, a Guest Editor for \textsc{IEEE Transactions on Cognitive Communications and Networking} Special Issue on “Machine Learning and Intelligent Signal Processing for Near-Field Technologies”, \textsc{IEEE Internet of Things Journal} Special Issue on “Next Generation Multiple Access for Internet-of-Things”, the NGMA workshop co-chairs of IEEE WCNC 2023, IEEE PIMRC 2023, and IEEE GLOBECOM 2023, and the “Mobile and Wireless Networks” symposium co-chair of IEEE GLOBECOM 2025. 
\end{IEEEbiography}

\begin{IEEEbiography}[{\includegraphics[width=1in, height=1.25in,clip, keepaspectratio]{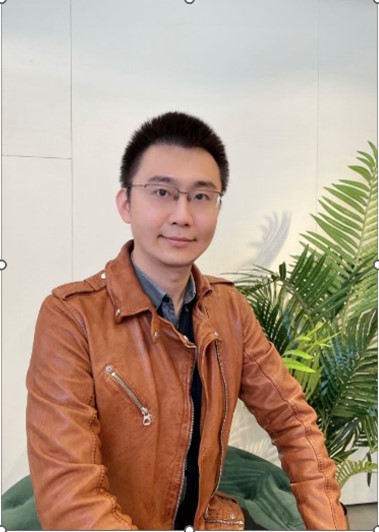}}]{Yuanwei Liu} (Fellow, IEEE, \url{http://www.eecs.qmul.ac.uk/~yuanwei})  has been a (tenured) full Professor in Department of Electrical and Electronic Engineering (EEE) at The University of Hong Kong (HKU) since September, 2024. Prior to that, he was a Senior Lecturer (Associate Professor) (2021-2024) and a Lecturer (Assistant Professor) (2017- 2021) at Queen Mary University of London (QMUL), London, U.K, and a Postdoctoral Research Fellow (2016-2017) at King's College London (KCL), London, U.K. He received the Ph.D. degree from QMUL in 2016.  His research interests include non-orthogonal multiple access, reconfigurable intelligent surface, near field communications, integrated sensing and communications, and machine learning. 

Yuanwei Liu is a Fellow of the IEEE, a Fellow of AAIA, a Web of Science Highly Cited Researcher, an IEEE Communication Society Distinguished Lecturer, an IEEE Vehicular Technology Society Distinguished Lecturer, the rapporteur of ETSI Industry Specification Group on Reconfigurable Intelligent Surfaces on work item of “Multi-functional Reconfigurable Intelligent Surfaces (RIS): Modelling, Optimisation, and Operation”, and the UK representative for the URSI Commission C on “Radio communication Systems and Signal Processing”. He was listed as one of 35 Innovators Under 35 China in 2022 by MIT Technology Review. He received IEEE ComSoc Outstanding Young Researcher Award for EMEA in 2020. He received the 2020 IEEE Signal Processing and Computing for Communications (SPCC) Technical Committee Early Achievement Award, IEEE Communication Theory Technical Committee (CTTC) 2021 Early Achievement Award. He received IEEE ComSoc Outstanding Nominee for Best Young Professionals Award in 2021. He is the co-recipient of the 2024 IEEE Communications Society Heinrich Hertz Award, the Best Student Paper Award in IEEE VTC2022-Fall, the Best Paper Award in ISWCS 2022, the 2022 IEEE SPCC-TC Best Paper Award, the 2023 IEEE ICCT Best Paper Award, and the 2023 IEEE ISAP Best Emerging Technologies Paper Award. He serves as the Co-Editor-in-Chief of IEEE ComSoc TC Newsletter, an Area Editor of IEEE Communications Letters, an Editor of IEEE Communications Surveys \& Tutorials, IEEE Transactions on Wireless Communications, IEEE Transactions on Vehicular Technology, IEEE Transactions on Network Science and Engineering, IEEE Transactions on Cognitive Communications and Networking, and IEEE Transactions on Communications (2018-2023). He serves as the (leading) Guest Editor for Proceedings of the IEEE on Next Generation Multiple Access, IEEE JSAC on Next Generation Multiple Access, IEEE JSTSP on Intelligent Signal Processing and Learning for Next Generation Multiple Access, and IEEE Network on Next Generation Multiple Access for 6G. He serves as the Publicity Co-Chair for IEEE VTC 2019-Fall, the Panel Co-Chair for IEEE WCNC 2024, Symposium Co-Chair for several flagship conferences such as IEEE GLOBECOM, ICC and VTC. He serves the academic Chair for the Next Generation Multiple Access Emerging Technology Initiative, vice chair of SPCC and Technical Committee on Cognitive Networks (TCCN).
\end{IEEEbiography}

\begin{IEEEbiography}[{\includegraphics[width=1in, height=1.25in,clip, keepaspectratio]{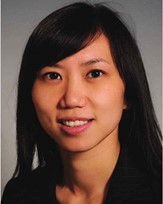}}] {Yue Chen} (Senior Member, IEEE) received the B.Eng. and M.Eng. degrees from the Beijing University of Posts and Telecommunications (BUPT), Beijing, China, in 1997 and 2000, respectively, and the Ph.D. degree from QMUL, London, U.K., in 2003.

She is currently a Professor of Telecommunications Engineering at the School of Electronic Engineering and Computer Science, Queen Mary University of London (QMUL), U.K. Her current research interests include wireless communication networks, mobile edge computing, smart energy systems, and the Internet of Things. 
\end{IEEEbiography}

\begin{IEEEbiography}[{\includegraphics[width=1in, height=1.25in,clip, keepaspectratio]{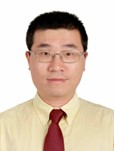}}]{Pan Zhiwen} (Member, IEEE) is currently a Professor with the National Mobile Communications Research Laboratory, Southeast University, Nanjing, China. During 2000–2001, he was involved in the research and standardization of 3G, and since 2002, he has been involved in the investigations on key technologies for IMT-A, 5G and 6G. He has authored or coauthored more than 50 papers recently, and holds more than 70 patents. His research interests include self-organizing networks, wireless networking, and radio transmission technology for wireless communications.
\end{IEEEbiography}

\end{document}